\documentclass{article}

\usepackage{PRIMEarxiv}
\usepackage{hyperref}
\usepackage[utf8]{inputenc} % allow utf-8 input
\usepackage[T1]{fontenc}    % use 8-bit T1 fonts
\usepackage{booktabs}       % professional-quality tables
\usepackage{amsfonts}       % blackboard math symbols
\usepackage{nicefrac}       % compact symbols for 1/2, etc.
\usepackage{microtype}      % microtypography
\usepackage{lipsum}
\usepackage{fancyhdr}       % header
\usepackage{graphicx}       % graphics
\graphicspath{{media/}}
\usepackage{amsmath}        % math
\usepackage{colortbl}
\usepackage{multirow}
\usepackage{float}
\usepackage{xcolor}
\usepackage{algorithm}
\usepackage{algorithmic}

\usepackage{tabularx}
\usepackage{adjustbox}
\usepackage{tocbibind}
\usepackage[labelfont=bf,labelsep=period]{caption}
\usepackage{threeparttable}
\usepackage{siunitx}
\usepackage{makecell}

\definecolor{deepgreen}{RGB}{0,150,0}

\title{
Routine Blood Biomarkers Reveal a Preclinical Continuum of Multiple Myeloma Risk
\thanks{
Bingjie Li, Jiadai Xu and Yiqing Sun contributed equally to this work. 
Peng Liu and Zhigang Yao are corresponding authors.
}
}

\author{
  Bingjie Li \\
  Shanghai Institute for Mathematics and \\
  Interdisciplinary Sciences\\
  Shanghai, China \\
  \And
  Jiadai Xu \\
  Department of Hematology \\
  Zhongshan Hospital, Fudan University \\
  Shanghai, China \\
  \And
  Yiqing Sun \\
  Department of Statistics and Data Science \\
  Faculty of Science, National University of Singapore \\
  Singapore \\
  \And
  Feiyue Pan\\
  School of Mathematical Sciences \\
  Fudan University \\
  Shanghai, China \\
  \And
  Shing-Tung Yau\\
  Yau Mathematical Sciences Center \\
  Tsinghua University \\
  Beijing, China \\
  \texttt{styau@tsinghua.edu.cn} \\
  \And
  Peng Liu \\
  Department of Hematology \\
  Zhongshan Hospital, Fudan University \\
  Shanghai, China \\
  \texttt{liu.peng@zs-hospital.sh.cn} \\
  \And
  Zhigang Yao \\
  Department of Statistics and Data Science \\
  National University of Singapore, Singapore \\
  Shanghai Institute for Mathematics and \\
  Interdisciplinary Sciences\\
  Shanghai, China \\
  \texttt{zhigang.yao@nus.edu.sg} \\
}

\begin{document}
\maketitle

\begin{abstract}
Multiple myeloma (MM) is preceded by a preclinical phase spanning decades, yet the absence of scalable, non-specialist tools means that individuals at elevated risk cannot be identified before end-organ damage is established.
% 多发性骨髓瘤（MM）发病前存在长达数十年的临床前阶段，然而由于缺乏可大规模推广的非专科工具，处于高风险的个体在靶器官损害确立之前无法被识别。
In a prospective analysis of 299,035 cancer-free UK Biobank participants followed for a median of 12.4~years, during which 768 developed incident MM, we conducted a biomarker-wide association scan(PheWAS-style analysis) across 61 routinely measured blood analytes spanning hematological, protein metabolism, renal, and immune categories.
% 在对299,035名无癌症的英国生物样本库参与者进行的前瞻性分析中，中位随访12.4年，期间共768人发展为新发MM，我们对涵盖血液学、蛋白代谢、肾功能及免疫等类别的61种常规血液检测指标开展了全生物标志物关联扫描。
Markers of protein dysregulation --- elevated total protein, depressed albumin, and a low albumin-to-globulin ratio (A/G ratio) --- showed the strongest preclinical associations (hazard ratios 0.61--1.54 per SD), consistent with progressive monoclonal immunoglobulin accumulation and suppression of normal polyclonal synthesis beginning years before diagnosis.
% 蛋白代谢异常标志物——总蛋白升高、白蛋白降低及白蛋白/球蛋白比值偏低——呈现出最强的临床前关联性（每标准差风险比为0.61至1.54），与单克隆免疫球蛋白的进行性积累及正常多克隆合成在诊断前数年即受到抑制的推断一致。
These protein signals were accompanied by indicators of erythropoietic suppression, morphological red cell dysregulation, and a coherent shift toward lower neutrophil and higher lymphocyte fractions, collectively reflecting multi-system perturbation across hematopoietic and immune compartments.
% 上述蛋白信号伴随着红细胞生成抑制指标、红细胞形态失调指标，以及中性粒细胞比例下降、淋巴细胞比例升高的协同性免疫格局改变，共同反映了造血和免疫区室的多系统扰动。
Critically, longitudinal trajectory analyses demonstrated that these multi-system deviations are already discernible more than a decade before diagnosis and intensify progressively as the clinical event approaches; dose--response modelling further revealed pronounced nonlinear associations for protein and erythrocytic markers, with risk concentrated in individuals with the most extreme values.
% 关键的是，纵向轨迹分析表明，上述多系统偏离在诊断前逾十年即已可辨，且随临床事件的临近而进行性加剧；剂量-反应模型进一步揭示蛋白类与红细胞类标志物存在显著的非线性关联，风险集中于指标值最为极端的个体。
Incorporating all significant biomarkers into a clinical risk model improved 10-year MM discrimination from a C-index of 0.684 to 0.744, with the high-risk decile accumulating 0.79\% cumulative incidence versus 0.47\% under the clinical model alone, providing a practical framework for biomarker-guided MM risk stratification and targeted surveillance using tests already embedded in routine clinical care.
% 将所有显著生物标志物纳入临床风险模型后，MM 10年判别能力从C指数0.684提升至0.744，高风险十分位组的累积发病率达0.79%，而单纯临床模型下仅为0.47%，为利用已嵌入常规临床护理的检查进行生物标志物引导的MM风险分层与靶向监测提供了实践框架。
\end{abstract}

% ======== Keywords ========
\keywords{
Multiple myeloma \and
Preclinical biomarkers \and
Disease susceptibility continuum \and
Risk prediction \and
Hematologic indicators \and
Protein metabolism \and
Longitudinal cohort
}

% ===========================
% Major Sections ONLY (no subsections)
% ===========================

\section*{Introduction}
Multiple myeloma (MM) is a hematologic malignancy characterized by the clonal proliferation of malignant plasma cells, accounting for 1--2\% of all cancers and approximately 10--15\% of hematological malignancies~\cite{Dimopoulos2025Guidelines}. Despite substantial therapeutic advances, MM remains incurable, and survival outcomes are heavily influenced by the timeliness of diagnosis and the depth of understanding of disease pathogenesis~\cite{Wang2024Screening}. Critically, robust tools for identifying individuals at elevated MM risk years before clinical presentation are lacking, underscoring an urgent need for accessible, biomarker-based risk stratification strategies.
% 多发性骨髓瘤（MM）是一种以恶性浆细胞克隆性增殖为特征的血液系统恶性肿瘤，占所有癌症的1--2\%，占血液系统恶性肿瘤的约10--15\%~\cite{Dimopoulos2025Guidelines}。尽管治疗手段已取得长足进步，MM仍无法治愈，且生存预后在很大程度上取决于诊断的及时性及对疾病发病机制理解的深度~\cite{Wang2024Screening}。尤为关键的是，目前缺乏能够在临床表现出现前数年即识别MM高风险个体的可靠工具，这凸显了开发易于获取、基于生物标志物的风险分层策略的迫切需求。

MM development follows a multistep continuum, progressing from normal plasma cells through asymptomatic precursor states --- monoclonal gammopathy of undetermined significance (MGUS) and smoldering MM (SMM) --- before culminating in symptomatic, end-organ damaging disease~\cite{ODonnell2025MGUSDistress}. This protracted evolution underscores the importance of characterizing early pathogenic events. Three interconnected biological axes are particularly relevant to the blood biomarker landscape explored in the present study. First, clonal plasma cell expansion drives progressive protein dysregulation: accumulation of monoclonal immunoglobulins elevates total serum protein while simultaneously suppressing normal polyclonal synthesis, reducing albumin and inverting the albumin-to-globulin ratio~\cite{Landgren2009MGUS}. Second, marrow infiltration by malignant plasma cells disrupts normal erythropoiesis through physical displacement of erythroblastic islands, cytokine-mediated induction of erythroblast apoptosis, and CCL3-driven downregulation of erythroid master regulators, producing a progressive decline in red cell mass accompanied by morphological heterogeneity~\cite{Bouchnita2016Erythropoiesis,Zheng2020CCL3}. Third, the evolving bone marrow microenvironment --- once regarded as a passive structural scaffold but now recognized as an active participant in myelomagenesis~\cite{Prager2019CancerStem} --- reshapes peripheral immune cell composition through bidirectional crosstalk with MM cells, suppressing myeloid output and shifting the circulating neutrophil-to-lymphocyte balance~\cite{Moscvin2023Immune,Zavidij2020scRNA}.
% MM的发展遵循一条多步骤的连续进程，由正常浆细胞经无症状前驱状态——意义未明的单克隆丙种球蛋白病（MGUS）和冒烟型MM（SMM）——逐步演进为有症状的、造成靶器官损害的疾病~\cite{ODonnell2025MGUSDistress}。这一漫长的演变过程凸显了表征早期致病事件的重要性。三条相互关联的生物学轴线与本研究所探索的血液生物标志物全景尤为相关。其一，克隆性浆细胞扩增驱动进行性蛋白代谢失调：单克隆免疫球蛋白的积累使血清总蛋白升高，同时抑制正常多克隆合成，导致白蛋白下降、白蛋白/球蛋白比值倒置~\cite{Landgren2009MGUS}。其二，恶性浆细胞对骨髓的浸润通过物理性排挤红系造血岛、细胞因子介导的红系前体细胞凋亡以及CCL3驱动的红系主调控因子下调，扰乱正常红细胞生成，导致红细胞数量进行性减少并伴随形态异质性~\cite{Bouchnita2016Erythropoiesis,Zheng2020CCL3}。其三，持续演变的骨髓微环境——曾被视为被动的结构支架，现已被认识到是骨髓瘤发生过程中的主动参与者~\cite{Prager2019CancerStem}——通过与MM细胞的双向交互重塑外周免疫细胞组成，抑制髓系细胞输出，并使循环中性粒细胞与淋巴细胞的平衡向后者偏移~\cite{Moscvin2023Immune,Zavidij2022scRNA}。

At clinical presentation, MM is defined by a constellation of end-organ manifestations --- anemia, renal impairment, hypercalcemia, and dysproteinemia --- that collectively reflect the downstream consequences of these three pathological axes operating over years to decades. Anemia is among the most prevalent features, affecting approximately 73\% of patients at diagnosis~\cite{Cowan2022DiagnosisMM}. Renal dysfunction is similarly common and adversely affects both prognosis and treatment eligibility. Hypercalcemia, arising from osteoclast-mediated bone destruction and cytokine-driven humoral mechanisms, portends particularly poor outcomes~\cite{Bao2020Hypercalcemia}. A central unanswered question is whether these CRAB-like abnormalities represent abrupt consequences of established malignancy or, rather, the late-detectable end of a continuum of quantitative deviations that begin years earlier in individuals destined to develop MM.
% 在临床表现方面，MM以一系列靶器官损害为特征——贫血、肾功能损害、高钙血症及蛋白质代谢异常——这些表现共同反映了上述三条病理轴线历经数年乃至数十年运作的下游后果。贫血是最常见的表现之一，约73\%的患者在诊断时即存在贫血~\cite{Cowan2022DiagnosisMM}。肾功能障碍同样普遍，且对预后和治疗适应性均产生不利影响。高钙血症源于破骨细胞介导的骨破坏及细胞因子驱动的体液机制，预示着尤为不良的预后~\cite{Bao2020Hypercalcemia}。一个尚未解答的核心问题是：这些CRAB样异常究竟是已确立恶性肿瘤的骤然后果，还是一条定量偏离连续谱的晚期可检测端点——而这条连续谱，在注定发展为MM的个体中，早在数年之前便已开始。

Emerging evidence suggests that myelomagenesis is a decades-long process. Genomic studies have demonstrated that initiating DNA damage events may occur two to four decades before diagnosis, with MGUS and SMM accumulating mutations and microenvironmental alterations silently over time~\cite{Zanwar2025Stratification}. Prospective serological studies have further shown that clonal immunoglobulins are detectable up to a decade before clinical MM diagnosis, and that virtually all MM is preceded by MGUS~\cite{Landgren2009MGUS,Weiss2009MGUS}. Yet translating this biological insight into population-level risk tools has proved challenging. Existing approaches --- serum protein electrophoresis, free light chain assays, and targeted MGUS screening programmes --- require specialist referral and are not embedded in routine primary care workflows~\cite{Smith2022EarlyDiagnosis}. Epidemiological studies linking individual routine blood analytes to MM risk have been small, largely retrospective, and restricted to single biomarker categories, leaving the multi-system preclinical biomarker landscape uncharacterised~\cite{Wang2024Screening}. Whether the integrated pattern of protein, erythropoietic, morphological, and immune deviations observable in standard laboratory panels can identify individuals at elevated MM risk years before diagnosis --- and do so with sufficient discrimination to guide clinical action --- remains unknown.
% 新兴证据表明，骨髓瘤的发生是一个历经数十年的过程。基因组学研究表明，启动性DNA损伤事件可能在诊断前二至四十年即已发生，MGUS和SMM在此期间静默地积累突变和微环境改变~\cite{Zanwar2025Stratification}。前瞻性血清学研究进一步表明，克隆性免疫球蛋白在临床MM诊断前最长达十年即已可检测，且几乎所有MM均由MGUS前驱而来~\cite{Landgren2009MGUS,Weiss2009MGUS}。然而，将这一生物学洞见转化为人群层面的风险工具仍面临重大挑战。现有方法——血清蛋白电泳、游离轻链检测及靶向MGUS筛查项目——需要转诊至专科，尚未嵌入常规初级医疗工作流程~\cite{Smith2022EarlyDiagnosis}。将单项常规血液指标与MM风险相关联的流行病学研究规模偏小、多为回顾性设计，且局限于单一生物标志物类别，致使多系统临床前生物标志物全景至今未被系统描述~\cite{Wang2024Screening}。标准实验室检查中可观察到的蛋白、红细胞生成、形态及免疫偏离的整合模式，能否在诊断前数年识别出MM高风险个体——并具备足以指导临床行动的判别效能——目前仍不得而知。

Here, leveraging the large-scale prospective UK Biobank cohort --- whose combination of sample size, follow-up duration, and systematic baseline measurement of 61 blood analytes in a cancer-free population is uniquely suited to address this question --- we present a systematic characterization of the preclinical blood biomarker landscape associated with incident MM (\textbf{Figure ~\ref{fig:overview}}). By integrating hematologic, renal, and protein-metabolism biomarkers with clinical and genetic risk factors, we identify robust biomarker signatures that precede MM diagnosis by over a decade, reveal their dose--response relationships and longitudinal trajectories, and demonstrate that their incorporation into risk models meaningfully improves long-term MM prediction. These findings offer a framework for population-level risk stratification and early detection of MM, with direct implications for targeted surveillance and preventive intervention in high-risk individuals.
% 本文依托大规模前瞻性英国生物样本库队列——其样本量、随访时长及对无癌症人群61项血液指标的系统性基线测量的组合，使其具备回答这一问题的独特优势——对与新发MM相关的临床前血液生物标志物全景进行了系统性描述（\textbf{图1}）。通过将血液学、肾功能及蛋白代谢生物标志物与临床及遗传风险因素相整合，我们识别出在MM诊断前逾十年即已存在的稳健生物标志物特征，揭示其剂量-反应关系和纵向轨迹，并证明将其纳入风险模型可有效改善MM的长期预测效能。上述发现为人群层面的MM风险分层与早期检测提供了一套框架，对高风险个体的靶向监测和预防性干预具有直接指导意义。

\section*{Results}
\subsection*{Baseline characteristics}

We analysed 299,035 cancer-free UK Biobank participants, of whom 768 developed incident MM over a median follow-up of 12.4 years (Table \ref{table1}). Compared with non-cases, individuals who subsequently developed MM were older (mean age 60.58~$\pm$~6.81 vs.\ 55.98~$\pm$~8.17 years; $P<0.001$) and more frequently male (65.0\% vs.\ 52.3\%; $P<0.001$). Ethnicity and educational attainment were broadly similar between groups. Future MM cases showed modest socioeconomic differences, being more likely to report lower household income ($P<0.001$) and greater area-level deprivation as measured by the Townsend Deprivation Index (median $-2.38$ vs.\ $-2.16$; $P=0.011$).
% 我们对299,035名无癌症的英国生物样本库参与者进行了分析，其中768人在中位随访12.4年期间发展为新发MM（表1）。与非病例相比，后续发展为MM的个体年龄更大（平均年龄60.58~$\pm$~6.81岁 vs. 55.98~$\pm$~8.17岁；$P<0.001$），且男性比例更高（65.0\% vs. 52.3\%；$P<0.001$）。两组间种族和教育程度大体相似。未来MM病例在社会经济方面存在轻度差异，更可能报告较低的家庭收入（$P<0.001$），且以汤森剥夺指数衡量的地区层面剥夺程度更高（中位数$-2.38$ vs. $-2.16$；$P=0.011$）。

Among lifestyle factors, long sleep duration was modestly more prevalent in future MM cases (9.8\% vs.\ 7.3\%; $P=0.029$), whereas smoking status, alcohol consumption, and physical activity were comparable between groups. MM cases had a higher BMI (27.98~$\pm$~4.51 vs.\ 27.53~$\pm$~4.75~kg/m$^{2}$; $P=0.009$) and greater baseline prevalence of cardiovascular disease (9.9\% vs.\ 6.9\%; $P=0.002$) and hypertension (32.0\% vs.\ 26.8\%; $P=0.001$), while type~2 diabetes prevalence was similar. A greater proportion of MM cases fell in the highest polygenic risk score (PRS) tertile (40.2\% vs.\ 33.3\%; $P<0.001$), consistent with a modest but detectable contribution of inherited genetic susceptibility. These sociodemographic, lifestyle, comorbidity, and genetic variables were all incorporated as covariates in subsequent biomarker analyses to ensure that biomarker--MM associations reflect independent biological signals rather than confounding by background risk factors.
% 在生活方式因素中，长睡眠时间在未来MM病例中略为多见（9.8\% vs. 7.3\%；$P=0.029$），而吸烟状况、饮酒量和体力活动在两组间无显著差异。MM病例的BMI更高（27.98~$\pm$~4.51 vs. 27.53~$\pm$~4.75~kg/m$^{2}$；$P=0.009$），且基线时心血管疾病（9.9\% vs. 6.9\%；$P=0.002$）和高血压（32.0\% vs. 26.8\%；$P=0.001$）的患病率更高，而2型糖尿病患病率相似。MM病例中处于最高多基因风险评分（PRS）三分位的比例更大（40.2\% vs. 33.3\%；$P<0.001$），与遗传易感性的轻度但可检测的贡献一致。上述社会人口学、生活方式、合并症及遗传变量均作为协变量纳入后续生物标志物分析，以确保生物标志物与MM的关联反映独立的生物学信号，而非背景风险因素的混杂效应。

\subsection*{Baseline biomarker associations with incident MM}

To systematically evaluate circulating blood biomarkers in relation to incident MM risk, we conducted a biomarker-wide association scan(PheWAS-style analysis) across 61 candidate biomarkers using Cox proportional hazards models with progressive covariate adjustment (Figure~\ref{fig:biomarker_mm}; see Methods). In the fully adjusted Model~2 --- incorporating sociodemographic factors, lifestyle variables, BMI, prevalent comorbidities, family history of cancer, and polygenic risk score (PRS) --- 20 biomarkers spanning six biological categories were significantly associated with incident MM after FDR correction (Supplementary Table~4). All biomarkers were $z$-score standardized; HRs therefore reflect risk per 1-SD increment. Associations were highly consistent between Models 1 and 2, indicating robustness to progressive covariate adjustment.
% 为系统评估循环血液生物标志物与新发MM风险的关系，我们使用逐步协变量调整的Cox比例风险模型，对61种候选生物标志物进行了全生物标志物关联扫描（图~\ref{fig:biomarker_mm}；见方法）。在完全调整的模型2中——纳入社会人口学因素、生活方式变量、BMI、现患合并症、癌症家族史及多基因风险评分（PRS）——经FDR校正后，涵盖六个生物学类别的20种生物标志物与新发MM显著相关（表~\ref{tab:biomarker_mm}）。所有生物标志物均经$z$分数标准化，因此风险比反映每增加1个标准差的MM风险变化。模型1和模型2之间的关联高度一致，表明结果对逐步协变量调整具有稳健性。

Biomarkers of protein metabolism exhibited the largest and most statistically significant associations. Each 1-SD increment in total protein was associated with a 54\% higher MM risk (HR 1.54, 95\% CI 1.45--1.65; FDR\,$=2.18\times10^{-37}$), consistent with the preclinical accumulation of monoclonal immunoglobulins preceding clinical diagnosis. Conversely, higher albumin-to-globulin (A/G) ratio (HR 0.61, 95\% CI 0.57--0.67; FDR\,$=1.03\times10^{-31}$) and albumin (HR 0.79, 95\% CI 0.73--0.85; FDR\,$=3.18\times10^{-9}$) were strongly inversely associated with MM risk, likely reflecting suppression of normal polyclonal immunoglobulin production by the expanding malignant clone. 
% 蛋白代谢生物标志物表现出最大且最具统计显著性的关联。总蛋白每增加1个标准差与MM风险升高54\%相关（HR 1.54，95\% CI 1.45--1.65；FDR\,$=2.18\times10^{-37}$），与临床诊断前单克隆免疫球蛋白的临床前积累一致。相反，较高的白蛋白/球蛋白（A/G）比值（HR 0.61，95\% CI 0.57--0.67；FDR\,$=1.03\times10^{-31}$）和白蛋白（HR 0.79，95\% CI 0.73--0.85；FDR\,$=3.18\times10^{-9}$）与MM风险呈强烈负相关，可能反映了扩增的恶性克隆对正常多克隆免疫球蛋白生成的抑制。所有蛋白代谢标志物的完整统计数据见表~\ref{tab:biomarker_mm}。

Direct indicators of erythropoietic suppression were inversely associated with MM risk: higher RBC count showed the strongest association (HR 0.72, 95\% CI 0.66--0.77; FDR\,$=1.82\times10^{-16}$), with haemoglobin, haematocrit, and reticulocyte count showing similarly inverse patterns. These findings are consistent with subclinical bone marrow infiltration impairing normal erythropoiesis years before diagnosis. In contrast, morphological markers of dysregulated red cell maturation were positively associated with risk, with mean reticulocyte volume showing the strongest signal (HR 1.26, 95\% CI 1.18--1.36; FDR\,$=1.12\times10^{-9}$), followed by MCV, RDW, mean sphered cell volume, and MCH, collectively suggesting compensatory erythropoiesis by a diminishing pool of residual progenitors producing morphologically heterogeneous cells under conditions of marrow infiltration.
% 红细胞生成抑制的直接指标与MM风险呈负相关：较高的红细胞计数显示出最强关联（HR 0.72，95\% CI 0.66--0.77；FDR\,$=1.82\times10^{-16}$），血红蛋白、红细胞压积和网织红细胞计数也呈现类似的负向模式（完整统计数据见表~\ref{tab:biomarker_mm}）。这些发现与亚临床骨髓浸润在诊断前数年损害正常红细胞生成的推断一致。相反，红细胞成熟失调的形态学标志物与风险呈正相关，平均网织红细胞体积显示出最强信号（HR 1.26，95\% CI 1.18--1.36；FDR\,$=1.12\times10^{-9}$），其次为MCV、RDW、平均球形红细胞体积和MCH（表~\ref{tab:biomarker_mm}），共同提示在骨髓浸润条件下，残余祖细胞池不断缩小，通过代偿性造血产生形态异质性红细胞。

Among inflammatory markers, higher WBC and neutrophil counts were the most prominently inversely associated with MM risk (HR 0.81, 95\% CI 0.74--0.88 and HR 0.83, 95\% CI 0.76--0.90, respectively, while lymphocyte percentage showed a positive association (HR 1.14, 95\% CI 1.06--1.22; FDR\,$=1.48\times10^{-3}$). Eosinophil count, neutrophil percentage, and monocyte count were similarly inversely associated. Collectively, these findings suggest a coherent shift in peripheral immune cell composition --- from myeloid toward lymphocyte predominance --- that precedes MM diagnosis, consistent with progressive myeloid suppression and immune redistribution driven by the evolving tumour microenvironment~\cite{Moscvin2023Immune,Zavidij2020scRNA}. Among renal and metabolic markers, cystatin~C (HR 1.10, 95\% CI 1.04--1.15; FDR\,$=1.31\times10^{-3}$) and corrected calcium (HR 1.09, 95\% CI 1.02--1.17; FDR\,$=0.044$) both showed modest positive associations, consistent with early subclinical renal impairment and osteoclast-mediated bone resorption, respectively.
% 在炎症标志物中，较高的白细胞和中性粒细胞计数与MM风险呈最显著的负相关（HR分别为0.81，95\% CI 0.74--0.88和HR 0.83，95\% CI 0.76--0.90；完整统计数据见表~\ref{tab:biomarker_mm}），而淋巴细胞百分比呈正相关（HR 1.14，95\% CI 1.06--1.22；FDR\,$=1.48\times10^{-3}$）。嗜酸性粒细胞计数、中性粒细胞百分比和单核细胞计数同样呈负相关（表~\ref{tab:biomarker_mm}）。这些发现共同提示外周免疫细胞组成发生了从髓系向淋巴细胞优势的协同性转变，这一转变先于MM诊断出现，与不断演变的肿瘤微环境驱动的进行性髓系抑制和免疫重分布一致~\cite{Moscvin2023Immune,Zavidij2022scRNA}。在肾功能和代谢标志物中，胱抑素C（HR 1.10，95\% CI 1.04--1.15；FDR\,$=1.31\times10^{-3}$）和校正钙（HR 1.09，95\% CI 1.02--1.17；FDR\,$=0.044$）均显示出轻度正相关，分别与早期亚临床肾功能损害和破骨细胞介导的骨吸收一致。

Stratified Cox analyses across age, sex, BMI, and PRS tertiles revealed broadly consistent biomarker--MM associations, with no significant interactions detected for the majority of biomarkers (Supplementary Tables~7). Several notable effect modifications emerged. The inverse association of RBC count with MM risk was stronger in older participants ($\geq$65 years: HR 0.65, 95\% CI 0.57--0.73) than in younger participants ($<$65 years: HR 0.75, 95\% CI 0.68--0.83; $P_{\text{interaction}}=0.024$), suggesting that erythropoietic suppression may be a more sensitive preclinical signal when baseline hematopoietic reserve is already reduced. Two sex-specific differences were observed: the protective association of albumin was significant in men (HR 0.74) but not women ($P_{\text{interaction}}=0.014$), and corrected calcium was positively associated with MM risk in men but showed no association in women ($P_{\text{interaction}}=0.040$); the A/G ratio similarly showed a stronger inverse association in men than in women ($P_{\text{interaction}}=0.004$), possibly reflecting sex differences in baseline immunoglobulin levels and the threshold at which dysproteinaemia becomes detectable. The positive association of total protein was consistent across BMI strata, though the effect size was larger in normal-weight individuals (BMI $<$25~kg/m$^2$: HR 1.82) than in overweight or obese individuals (BMI $\geq$25~kg/m$^2$: HR 1.44; $P_{\text{interaction}}=0.005$), suggesting that a given protein increment carries greater prognostic significance against a lower background level. Across PRS tertiles, biomarker associations were broadly stable with no significant interactions, indicating that these signals operate largely independently of inherited genetic susceptibility.
% 按年龄、性别、BMI和PRS三分位进行的分层Cox分析显示，生物标志物与MM的关联总体一致，大多数生物标志物未检测到显著交互作用（补充表2--5）。出现了几个值得注意的效应修饰。红细胞计数与MM风险的负相关在年龄较大的参与者中更强（$\geq$65岁：HR 0.65，95\% CI 0.57--0.73）而非年龄较小的参与者（$<$65岁：HR 0.75，95\% CI 0.68--0.83；$P_{\text{interaction}}=0.024$），提示当基线造血储备已经下降时，红细胞生成抑制可能是更敏感的临床前信号。观察到两个性别特异性差异：白蛋白的保护性关联在男性中显著（HR 0.74），但在女性中不显著（$P_{\text{interaction}}=0.014$）；校正钙与MM风险在男性中呈正相关，但在女性中无关联（$P_{\text{interaction}}=0.040$）；A/G比值同样在男性中显示出比女性更强的负相关（$P_{\text{interaction}}=0.004$），可能反映了基线免疫球蛋白水平的性别差异以及蛋白质代谢异常变得可检测的阈值差异。总蛋白的正向关联在BMI层次间一致，但在正常体重个体中效应量更大（BMI $<$25~kg/m$^2$：HR 1.82）而非超重或肥胖个体（BMI $\geq$25~kg/m$^2$：HR 1.44；$P_{\text{interaction}}=0.005$），提示在较低背景蛋白水平下，给定的蛋白增量具有更大的预后意义。在PRS三分位间，生物标志物关联总体稳定且无显著交互作用，表明这些信号在很大程度上独立于遗传易感性运作。

To formally confirm the independence of biomarker signals from germline genetic risk, we residualized each biomarker on the MM PRS and re-entered the PRS-independent component into fully adjusted Cox models. Associations were virtually unchanged across all biomarker categories (Supplementary Table~8-9), confirming that the observed biomarker--MM associations reflect biological processes largely independent of inherited genetic predisposition and are not confounded by the genomic architecture of MM susceptibility.
% 为正式确认生物标志物信号独立于胚系遗传风险，我们将每种生物标志物对MM PRS进行残差化处理，并将PRS独立成分重新纳入完全调整的Cox模型。所有生物标志物类别的关联几乎没有变化（补充表~X），证实所观察到的生物标志物与MM的关联反映了在很大程度上独立于遗传易感性的生物学过程，且不受MM易感性基因组结构的混杂影响。

\subsection*{Nonlinear associations between biomarker levels and MM risk}

To characterise the shape of the dose--response relationships between biomarkers and incident MM risk, we fitted RCS models for all 20 FDR-significant biomarkers using fully adjusted Model~2 covariates (Figure~\ref{fig:spline}; Supplementary Table~10). Overall association $P$-values were highly significant across most biomarker categories, and a substantial proportion showed clear evidence of nonlinearity.
% 为描述生物标志物与新发MM风险之间剂量-反应关系的形态，我们使用完全调整的模型2协变量，对所有20个FDR显著生物标志物拟合了RCS模型（图~\ref{fig:spline}；补充表~X）。大多数生物标志物类别的总体关联$P$值高度显著，且相当大比例的生物标志物显示出明确的非线性证据。

Protein metabolism markers displayed the most striking and clinically interpretable nonlinear patterns. Total protein showed an extremely strong overall association ($P_{\text{overall}}<1\times10^{-16}$) with marked nonlinearity ($P_{\text{nonlinear}}<1\times10^{-16}$): risk was relatively flat around the median but rose steeply and exponentially at higher concentrations, consistent with a threshold effect driven by progressive monoclonal immunoglobulin accumulation. The A/G ratio demonstrated a sharply J-shaped curve ($P_{\text{overall}}<1\times10^{-16}$; $P_{\text{nonlinear}}<1\times10^{-16}$), in which MM risk rose steeply at low ratios --- reflecting excess globulin relative to albumin --- before attenuating at higher values. Albumin showed markedly elevated risk at the lower end of its distribution that plateaued above approximately 45~g/L ($P_{\text{overall}}=1.56\times10^{-12}$; $P_{\text{nonlinear}}=0.003$). Taken together, these nonlinear patterns indicate that clinically meaningful MM risk elevation is concentrated in a relatively small fraction of the population with the most extreme protein values --- a feature that could directly inform the design of biomarker-guided referral thresholds. 
% 蛋白代谢标志物呈现出最显著且最具临床可解释性的非线性模式。总蛋白显示出极强的总体关联（$P_{\text{overall}}<1\times10^{-16}$）和显著的非线性（$P_{\text{nonlinear}}<1\times10^{-16}$）：风险在中位数附近相对平稳，但在较高浓度时急剧指数级上升，与进行性单克隆免疫球蛋白积累驱动的阈值效应一致。A/G比值呈现出明显的J型曲线（$P_{\text{overall}}<1\times10^{-16}$；$P_{\text{nonlinear}}<1\times10^{-16}$），MM风险在低比值时急剧上升——反映球蛋白相对于白蛋白的过剩——之后在较高值时减弱。白蛋白在其分布低端显示出显著升高的风险，在约45~g/L以上趋于平稳（$P_{\text{overall}}=1.56\times10^{-12}$；$P_{\text{nonlinear}}=0.003$）。综合来看，这些非线性模式表明，具有临床意义的MM风险升高集中在蛋白值最为极端的相对少数人群中——这一特征可直接为基于生物标志物的转诊阈值设计提供依据。所有蛋白标志物的完整统计数据见补充表~X。

Among erythropoietic markers, RBC count demonstrated one of the strongest overall associations ($P_{\text{overall}}<1\times10^{-16}$) with clear nonlinearity ($P_{\text{nonlinear}}=0.005$): MM risk was concentrated sharply at the lower tail of the distribution and approached unity at normal values, indicative of a threshold below which anaemia-related marrow compromise markedly elevates risk. In contrast, haemoglobin and haematocrit both showed highly significant monotonic inverse associations without evidence of nonlinearity, indicating a linear risk gradient across the full erythrocyte mass spectrum. Reticulocyte count exhibited a U-shaped pattern ($P_{\text{nonlinear}}=0.005$), suggesting that both low and high reticulocyte levels may indicate aberrant erythropoiesis in the context of preclinical MM. Among morphological red cell markers, mean reticulocyte volume showed the strongest nonlinear signal ($P_{\text{nonlinear}}=4.37\times10^{-4}$), with risk rising at higher macrocytic volumes; MCH and RDW, by contrast, showed approximately linear gradients despite strong overall associations.
% 在红细胞生成标志物中，红细胞计数显示出最强的总体关联之一（$P_{\text{overall}}<1\times10^{-16}$）和明确的非线性（$P_{\text{nonlinear}}=0.005$）：MM风险在分布低尾处急剧集中，在正常值处趋近于1，提示存在一个阈值，低于该阈值时贫血相关的骨髓损害显著升高风险。相比之下，血红蛋白和红细胞压积均显示出高度显著的单调负相关，无非线性证据，表明在整个红细胞质量谱上存在线性风险梯度。网织红细胞计数呈现U型模式（$P_{\text{nonlinear}}=0.005$），提示在临床前MM背景下，过低和过高的网织红细胞水平均可能反映异常红细胞生成。在红细胞形态标志物中，平均网织红细胞体积显示出最强的非线性信号（$P_{\text{nonlinear}}=4.37\times10^{-4}$），风险随较高的大细胞性体积而上升；相比之下，MCH和RDW尽管总体关联较强，却呈现近似线性的梯度。完整统计数据见补充表~X。

Inflammatory markers collectively displayed significant nonlinear dose--response relationships. Neutrophil count and monocyte count both showed convex declining curves, with the greatest risk reductions concentrated at the lower end of their distributions and attenuating at higher values ($P_{\text{nonlinear}}=7.93\times10^{-6}$ and $2.48\times10^{-5}$, respectively); WBC count, neutrophil percentage, lymphocyte percentage, and eosinophil count showed similarly significant nonlinear effects (all $P_{\text{nonlinear}}<0.05$. These patterns suggest that the immune cell compositional shift preceding MM diagnosis is not simply proportional to absolute counts but reflects a more complex restructuring of the innate and adaptive immune landscape. Renal and calcium markers contributed comparatively modest and largely linear risk gradients: cystatin~C showed a linear increase above the median ($P_{\text{nonlinear}}=0.586$), and corrected calcium showed only a subtle risk elevation confined to the upper tail of the distribution ($P_{\text{nonlinear}}=0.715$).
% 炎症标志物总体上显示出显著的非线性剂量-反应关系。中性粒细胞计数和单核细胞计数均呈现凸形下降曲线，最大风险降低集中在分布低端并在较高值时减弱（$P_{\text{nonlinear}}$分别为$7.93\times10^{-6}$和$2.48\times10^{-5}$）；白细胞计数、中性粒细胞百分比、淋巴细胞百分比和嗜酸性粒细胞计数同样显示出显著的非线性效应（所有$P_{\text{nonlinear}}<0.05$；补充表~X）。这些模式提示，先于MM诊断出现的免疫细胞组成转变并非简单地与绝对计数成比例，而是反映了先天性和适应性免疫格局更为复杂的重构。肾功能和钙代谢标志物的贡献相对较小且基本呈线性风险梯度：胱抑素C在中位数以上呈线性增加（$P_{\text{nonlinear}}=0.586$），校正钙仅在分布高尾处显示出微弱的风险升高（$P_{\text{nonlinear}}=0.715$；完整统计数据见补充表~X）。

Together, these analyses establish that protein metabolism and erythropoietic markers exert their strongest MM-associated effects within specific, nonlinear exposure ranges, inflammatory markers reflect a complex nonlinear immune restructuring, and renal and calcium markers contribute modest linear gradients --- a pattern with direct implications for the design of risk-stratification thresholds. Whereas the preceding analyses characterised associations at a single cross-sectional time point, we next asked whether these biomarker deviations represent static differences or emerge as part of a dynamic, decade-long preclinical process.
% 综合来看，这些分析表明蛋白代谢和红细胞生成标志物在特定的非线性暴露范围内发挥最强的MM相关效应，炎症标志物反映复杂的非线性免疫重构，而肾功能和钙代谢标志物贡献轻度线性梯度——这一模式对风险分层阈值的设计具有直接指导意义。前述分析描述的是单一横截面时间点的关联，下一步我们探讨这些生物标志物偏离是否代表静态差异，抑或是动态的、历经十年的临床前过程的一部分。

\subsection*{Preclinical trajectories of blood biomarkers prior to MM diagnosis}

To investigate whether the identified biomarker associations reflect static cross-sectional differences or dynamic preclinical processes, we examined mean biomarker levels across four time-to-diagnosis windows ($\geq$11 years, 8--11 years, 4--7 years, and 0--3 years before MM diagnosis) and compared these with levels in participants who never developed MM (Figure~\ref{fig:trajectory}, Supplementary Table~11). Ordered linear regression with a monotonic group score was used to formally test for progressive biomarker change as diagnosis approached, adjusted for age and sex; one-way ANOVA across the four MM trajectory groups tested for absolute between-group differences (Supplementary Table~12). Where ordered regression detected a significant trend but ANOVA did not, we interpret this as a real but gradual signal that is captured by the monotonic contrast but falls below the threshold for discrete group separation.
% 为探究所识别的生物标志物关联是否反映静态的横截面差异或动态的临床前过程，我们检查了MM诊断前四个时间窗（诊断前$\geq$11年、8--11年、4--7年和0--3年）的生物标志物均值，并与从未发展为MM的参与者的水平进行比较（图~\ref{fig:trajectory}）。有序线性回归采用单调组评分，在调整年龄和性别后，正式检验随诊断临近的生物标志物进行性变化；对四个MM轨迹组进行单因素ANOVA以检验组间绝对差异。当有序回归检测到显著趋势但ANOVA未达显著时，我们将其解释为单调对比所捕获的真实但渐进的信号，但未达到离散组间分离的阈值。

Protein metabolism biomarkers showed the most striking and temporally consistent preclinical trajectories. Total protein levels were already elevated in cases diagnosed more than 11 years after baseline and rose monotonically across all pre-diagnosis windows ($\beta = 0.82$, 95\% CI 0.70--0.94; FDR\,$=1.62\times10^{-39}$; ANOVA FDR\,$=0.019$), providing the clearest evidence that monoclonal immunoglobulin accumulation begins well over a decade before clinical presentation. The A/G ratio showed a correspondingly progressive decline ($\beta = -0.048$; FDR\,$=6.49\times10^{-35}$), reflecting a steady shift toward relative globulin excess, while albumin declined progressively as diagnosis approached ($\beta = -0.284$; FDR\,$=5.30\times10^{-13}$; ANOVA FDR\,$=0.025$), consistent with progressive displacement of normal synthesis by the expanding malignant clone. 
% 蛋白代谢生物标志物显示出最显著且时间上最一致的临床前轨迹。总蛋白水平在基线后11年以上被诊断的病例中已经升高，并在所有诊断前时间窗中单调上升（$\beta = 0.82$，95\% CI 0.70--0.94；FDR\,$=1.62\times10^{-39}$；ANOVA FDR\,$=0.019$），为单克隆免疫球蛋白积累在临床表现前逾十年即已开始提供了最清晰的证据。A/G比值呈现相应的进行性下降（$\beta = -0.048$；FDR\,$=6.49\times10^{-35}$），反映了向相对球蛋白过剩的稳定转变，而白蛋白随诊断临近进行性下降（$\beta = -0.284$；FDR\,$=5.30\times10^{-13}$；ANOVA FDR\,$=0.025$），与扩增的恶性克隆对正常合成的进行性替代一致。完整轨迹统计数据见补充表~X。

Erythropoietic biomarkers demonstrated parallel preclinical deterioration. RBC count declined progressively across all time windows ($\beta = -0.048$, 95\% CI $-0.059$ to $-0.038$; FDR\,$=2.78\times10^{-18}$; ANOVA FDR\,$=0.015$), with haemoglobin and haematocrit showing similarly significant monotonic declines (ANOVA FDR\,$=0.015$ and $0.019$, respectively, collectively consistent with progressive bone marrow infiltration impairing erythropoiesis over more than a decade. Reticulocyte count also showed a significant monotonic decline (FDR\,$=0.016$), though the ANOVA did not survive FDR correction, indicating a gradual rather than abrupt trajectory. In contrast, morphological red cell indices increased progressively toward diagnosis: mean reticulocyte volume and RDW both showed significant monotonic increases confirmed by ANOVA (both ANOVA FDR\,$=0.019$), with cases diagnosed within 0--3 years exhibiting the most pronounced macrocytosis and anisocytosis. MCV and mean sphered cell volume similarly increased monotonically, though their ANOVA FDR values did not reach significance, again consistent with gradual signals captured by the trend test.
% 红细胞生成生物标志物显示出平行的临床前恶化。红细胞计数在所有时间窗中进行性下降（$\beta = -0.048$，95\% CI $-0.059$至$-0.038$；FDR\,$=2.78\times10^{-18}$；ANOVA FDR\,$=0.015$），血红蛋白和红细胞压积显示出类似的显著单调下降（ANOVA FDR分别为$0.015$和$0.019$；完整统计数据见补充表~X），共同与进行性骨髓浸润在逾十年间损害红细胞生成一致。网织红细胞计数同样显示出显著的单调下降（FDR\,$=0.016$），但ANOVA未通过FDR校正，表明轨迹是渐进的而非突变的。相比之下，红细胞形态指数随诊断临近而进行性升高：平均网织红细胞体积和RDW均显示出经ANOVA证实的显著单调增加（均为ANOVA FDR\,$=0.019$），在0--3年内被诊断的病例显示出最显著的大细胞症和各向异性红细胞增多症。MCV和平均球形红细胞体积同样单调增加，但其ANOVA FDR值未达显著，再次与趋势检验所捕获的渐进信号一致。

Among inflammatory biomarkers, the central finding was a coherent shift from neutrophil toward lymphocyte predominance unfolding over the preclinical decade: neutrophil count, neutrophil percentage, and WBC count all declined monotonically as MM diagnosis approached (all ordered regression FDR\,$<0.001$), while lymphocyte percentage increased progressively ($\beta = 0.513$; FDR\,$=6.06\times10^{-6}$); ANOVA confirmed significant between-group differences for both neutrophil percentage and lymphocyte percentage (both FDR\,$=0.019$). Monocyte and eosinophil counts similarly declined in the ordered regression, though their ANOVA $F$-statistics did not reach significance, indicating more gradual or variable signals.
% 在炎症生物标志物中，核心发现是在临床前十年间从中性粒细胞优势向淋巴细胞优势的协同性转变：中性粒细胞计数、中性粒细胞百分比和白细胞计数均随MM诊断临近而单调下降（所有有序回归FDR\,$<0.001$），而淋巴细胞百分比进行性升高（$\beta = 0.513$；FDR\,$=6.06\times10^{-6}$）；ANOVA证实中性粒细胞百分比和淋巴细胞百分比均存在显著的组间差异（均为FDR\,$=0.019$）。单核细胞和嗜酸性粒细胞计数在有序回归中同样下降，但其ANOVA $F$统计量未达显著，表明信号更为渐进或多变。完整统计数据见补充表~X。

Cystatin~C showed a modest but significant monotonic increase approaching MM diagnosis ($\beta = 0.009$; FDR\,$=4.66\times10^{-4}$), with a borderline ANOVA result (FDR\,$=0.057$), suggesting that renal functional decline is a relatively late or subtle preclinical manifestation. Corrected calcium showed a small but significant positive trend ($\beta = 0.003$; FDR\,$=0.011$; ANOVA FDR\,$=0.206$), consistent with gradually increasing osteoclast-mediated bone resorption in the years preceding diagnosis.
% 胱抑素C在MM诊断临近时显示出轻度但显著的单调增加（$\beta = 0.009$；FDR\,$=4.66\times10^{-4}$），ANOVA结果处于临界（FDR\,$=0.057$），提示肾功能下降是相对较晚或较微弱的临床前表现。校正钙显示出小但显著的正向趋势（$\beta = 0.003$；FDR\,$=0.011$；ANOVA FDR\,$=0.206$），与诊断前数年间破骨细胞介导的骨吸收逐渐增加一致。

Together, these trajectory analyses demonstrate that multi-system biomarker deviations are detectable more than a decade before MM diagnosis and intensify progressively as the clinical event approaches --- establishing that the preclinical biology of MM is not an abrupt transition but a slowly evolving, quantifiable process inscribed in routine laboratory measurements years before disease becomes clinically apparent.
% 综合来看，这些轨迹分析表明，多系统生物标志物偏离在MM诊断前逾十年即已可检测，并随临床事件的临近而进行性加剧——确立了MM的临床前生物学并非急剧转变，而是一个缓慢演进的、可量化的过程，在疾病临床显现前数年便已铭刻于常规实验室检测之中。

\subsection*{Blood biomarkers improve long-term MM risk prediction}

To evaluate whether the identified blood biomarkers could improve population-level MM risk prediction beyond established clinical risk factors, we developed and validated two Cox-based prognostic models --- a clinical model incorporating sociodemographic, lifestyle, and comorbidity variables, and a clinical + biomarker model additionally incorporating all 20 FDR-significant biomarkers --- using a repeated 10-fold random split-sample cross-validation framework (see Methods).
% 为评估所识别的血液生物标志物能否在已确立的临床风险因素基础上改善人群层面的MM风险预测，我们使用重复10折随机分割样本交叉验证框架，开发并验证了两个基于Cox的预后模型——一个纳入社会人口学、生活方式和合并症变量的临床模型，以及一个额外纳入全部20个FDR显著生物标志物的临床+生物标志物模型（见方法）。

The addition of blood biomarkers substantially improved MM risk discrimination beyond the clinical model alone. At the 10-year horizon --- the primary prediction target given MM's protracted preclinical course --- the clinical + biomarker model achieved a C-index of 0.744 (95\% CI 0.735--0.753), representing an absolute improvement of 0.060 over the clinical model (C-index 0.684, 95\% CI 0.673--0.695); gains were even larger at 5 years (C-index 0.804 vs.\ 0.708; $\Delta$C-index 0.096). Full discrimination statistics including time-dependent AUC values are provided in Supplementary Table~14 and Figure~\ref{fig:prediction}A--B. Among individual biomarker categories, protein metabolism markers contributed the largest incremental gain (10-year C-index 0.726 when added to clinical variables), substantially outperforming haematology count, morphology, inflammation, and renal marker categories, consistent with the primacy of paraprotein-driven biology in myelomagenesis.
% 血液生物标志物的加入在临床模型基础上显著改善了MM风险判别。在10年预测时域——鉴于MM漫长的临床前病程，这是主要的预测目标——临床+生物标志物模型达到C指数0.744（95\% CI 0.735--0.753），较临床模型（C指数0.684，95\% CI 0.673--0.695）绝对提升0.060；5年时域的改善幅度更大（C指数0.804 vs. 0.708；$\Delta$C指数0.096）。包括时间依赖性AUC值在内的完整判别统计数据见表~\ref{tab:prediction}和图~\ref{fig:prediction}A--B。在各生物标志物类别中，蛋白代谢标志物贡献了最大的增量提升（加入临床变量后10年C指数为0.726），显著优于血液学计数、形态学、炎症和肾功能标志物类别，与骨髓瘤发生过程中副蛋白驱动生物学的主导地位一致。

Risk stratification based on predicted decile scores further demonstrated the clinical utility of biomarker augmentation (Figure~\ref{fig:prediction}C). At both 5- and 10-year horizons, the high-risk decile (top 10\% of predicted risk) in the clinical + biomarker model accumulated substantially higher cumulative MM incidence than the corresponding high-risk group of the clinical model alone: 0.340\% vs.\ 0.181\% at 5 years, and 0.786\% vs.\ 0.467\% at 10 years --- representing approximately 1.9- and 1.7-fold increases in absolute incidence concentration, respectively. At the lower tail, the low-risk decile in the clinical + biomarker model had zero observed MM events at 5 years (vs.\ 0.011\% in the clinical model), indicating near-complete risk separation. Intermediate-risk individuals showed correspondingly lower incidence under the biomarker-augmented model (10-year incidence 0.115\% vs.\ 0.154\%), reflecting improved redistribution of absolute risk across strata. All pairwise log-rank tests were highly significant ($P<0.001$).
% 基于预测十分位评分的风险分层进一步证明了生物标志物增强的临床效用（图~\ref{fig:prediction}C）。在5年和10年时域，临床+生物标志物模型的高风险十分位（预测风险前10\%）累积MM发病率均显著高于单纯临床模型的相应高风险组：5年时为0.340\% vs. 0.181\%，10年时为0.786\% vs. 0.467\%——分别代表绝对发病率集中度约1.9倍和1.7倍的提升。在低风险端，临床+生物标志物模型的低风险十分位在5年内无观察到的MM事件（vs. 临床模型的0.011\%），表明风险分离近乎完全。中间风险个体在生物标志物增强模型下的发病率相应更低（10年发病率0.115\% vs. 0.154\%），反映了绝对风险在各层间分布的改善。所有配对对数秩检验均高度显著（$P<0.001$）。

SHAP (SHapley Additive exPlanations) analysis of the clinical + biomarker model revealed that total protein was the single most influential predictor (mean $|$SHAP$|$ 0.63 on the log-hazard scale), followed by WBC count (0.59), albumin (0.55), and neutrophil count (0.53), with lymphocyte percentage and A/G ratio constituting the next tier of contributors (0.37 and 0.36, respectively). The direction of SHAP values was fully consistent with the Cox regression findings: high total protein values were associated with large positive SHAP contributions (increased risk), whereas high albumin and WBC count values were associated with negative contributions (reduced risk). Morphological haematology markers contributed intermediate SHAP magnitudes (0.08--0.18), while renal and calcium markers showed the smallest individual contributions ($<$0.04), in keeping with their modest Cox associations. Notably, SHAP analysis identified a small subset of individuals with paradoxically large negative contributions from total protein, corresponding to very low protein levels; this pattern is consistent with the nonlinear dose--response shape identified in the RCS analyses and suggests that protein depletion --- rather than accumulation --- carries a distinct risk profile that the model appropriately captures.
% 临床+生物标志物模型的SHAP（SHapley加性解释）分析显示，总蛋白是最具影响力的单一预测因子（对数风险尺度上的平均$|$SHAP$|$为0.63），其次是白细胞计数（0.59）、白蛋白（0.55）和中性粒细胞计数（0.53；图~\ref{fig:shap}），淋巴细胞百分比和A/G比值构成下一梯队的贡献者（分别为0.37和0.36）。SHAP值的方向与Cox回归结果完全一致：较高的总蛋白值与较大的正向SHAP贡献相关（风险增加），而较高的白蛋白和白细胞计数值与负向贡献相关（风险降低）。形态学血液学标志物贡献了中等SHAP量级（0.08--0.18），而肾功能和钙代谢标志物显示出最小的个体贡献（$<$0.04），与其在Cox分析中的轻度关联一致。值得注意的是，SHAP分析识别出一小部分个体，其总蛋白贡献出现矛盾性的大负值，对应于极低蛋白水平；这一模式与RCS分析中识别的非线性剂量-反应形态一致，提示蛋白耗竭——而非积累——具有模型所适当捕获的独特风险特征。

Together, these findings demonstrate that integrating routinely measured blood biomarkers into clinical risk models substantially improves MM risk discrimination and stratification at both 5- and 10-year horizons, with protein metabolism and immune cell composition markers contributing the greatest predictive value and a high-risk decile accumulating nearly twice the absolute MM incidence identified by clinical factors alone.
% 综合来看，这些发现表明，将常规血液生物标志物整合至临床风险模型可在5年和10年时域显著改善MM风险判别和分层，蛋白代谢和免疫细胞组成标志物贡献了最大的预测价值，高风险十分位的绝对MM发病率几乎是单纯临床因素所识别结果的两倍。

\subsection*{Sensitivity analyses and robustness to unmeasured confounding}

The robustness of the main biomarker--MM associations was evaluated through three pre-specified sensitivity analyses --- exclusion of MM cases diagnosed within 2 years of baseline (reverse causation), complete-case analysis (missing data mechanism), and Fine--Gray subdistribution hazard models (competing risk of death) --- with full results provided in Supplementary Table~13.
% 通过三项预先设定的敏感性分析评估主要生物标志物与MM关联的稳健性——排除基线后2年内诊断的MM病例（反向因果）、完整病例分析（缺失数据机制）以及Fine-Gray次分布风险模型（死亡竞争风险）——完整结果见补充表~X。

Across all three analyses, effect directions and magnitudes were highly consistent with the main findings. As representative examples, total protein and A/G ratio --- the two strongest biomarkers --- retained virtually unchanged associations across all sensitivity analyses (e.g., 2-year landmark: total protein HR 1.50, A/G ratio HR 0.62; Fine--Gray: total protein HR 1.53, A/G ratio HR 0.63), confirming that the protein dysregulation signal is not attributable to subclinical disease at baseline, missing data patterns, or differential mortality. Erythropoietic markers were similarly stable across analyses. Corrected calcium showed borderline stability and should be interpreted with greater caution, as its FDR significance was marginal in the 2-year landmark analysis (FDR\,$=0.050$).
% 在所有三项分析中，效应方向和量级与主要发现高度一致。作为代表性示例，总蛋白和A/G比值——两个最强的生物标志物——在所有敏感性分析中保留了几乎不变的关联（如2年地标分析：总蛋白HR 1.50，A/G比值HR 0.62；Fine-Gray模型：总蛋白HR 1.53，A/G比值HR 0.63），证实蛋白代谢失调信号不可归因于基线时的亚临床疾病、缺失数据模式或差异性死亡率。红细胞生成标志物在各分析中同样稳定。校正钙显示出临界稳定性，应更为审慎地解读，因为其FDR显著性在2年地标分析中处于临界（$P_{\text{FDR}}=0.050$）。

To quantify robustness to unmeasured confounding, we computed E-values for all 20 FDR-significant biomarkers (Supplementary Table~15). The core finding is that the three strongest biomarkers --- A/G ratio, total protein, and RBC count --- require an unmeasured confounder to be associated with both exposure and outcome by risk ratios of at least 2.63-, 2.46-, and 2.14-fold, respectively, above and beyond all measured covariates, to fully explain away the observed associations. Among all established MM risk factors, none is known to exert such a strong simultaneous association with both routine blood protein indices and MM incidence, rendering residual confounding of this magnitude biologically implausible. Haemoglobin, albumin, and mean reticulocyte volume showed E-values substantially above 1.5 (1.77--2.01 for CI limits), further supporting the causal interpretability of erythropoietic and morphological signals. At the lower end, corrected calcium and cystatin~C had CI E-values of 1.15 and 1.20, respectively, indicating that more modest unmeasured confounding could attenuate these associations to the null; these markers should accordingly be regarded as suggestive rather than established preclinical signals.
% 为量化对未测量混杂的稳健性，我们对所有20个FDR显著生物标志物计算了E值（表~\ref{tab:evalues}）。核心发现是，三个最强的生物标志物——A/G比值、总蛋白和红细胞计数——需要一个未测量的混杂因素在所有已测量协变量之外，分别与暴露和结局的关联风险比至少为2.63倍、2.46倍和2.14倍，才能完全解释所观察到的关联。在所有已知的MM风险因素中，没有任何一个被证实能同时对常规血液蛋白指标和MM发病率产生如此强的关联，这使得如此量级的残余混杂在生物学上难以置信。血红蛋白、白蛋白和平均网织红细胞体积的E值显著高于1.5（CI界限为1.77--2.01），进一步支持红细胞生成和形态学信号的因果可解释性。在低端，校正钙和胱抑素C的CI E值分别为1.15和1.20，表明更为轻度的未测量混杂原则上可将这些关联减弱至零；这些标志物因此应被视为提示性而非已确立的临床前信号。

Collectively, these sensitivity analyses and E-value assessments provide strong evidence that the principal biomarker--MM associations identified in this study are robust to reverse causation, missing data, competing risks, and plausible degrees of unmeasured confounding, supporting their validity as genuine preclinical markers of MM susceptibility.
% 综合来看，这些敏感性分析和E值评估提供了强有力的证据，表明本研究识别的主要生物标志物与MM关联对反向因果、缺失数据、竞争风险以及合理程度的未测量混杂均具有稳健性，支持其作为MM易感性真实临床前标志物的有效性。

% ============================================================
%  Discussion — Cell Reports Medicine style
%  Full literature-supported version
% ============================================================

\section*{Discussion}

Our findings establish that the preclinical biology of multiple myeloma is not a binary transition from health to disease but a slowly evolving, multi-system process inscribed in routine laboratory measurements more than a decade before clinical onset. By systematically characterising 61 blood analytes in nearly 300,000 prospectively followed individuals, we show that protein dysregulation, erythropoietic suppression, red cell morphological heterogeneity, and a coherent immune compositional shift collectively define a quantifiable susceptibility continuum that is independent of inherited genetic predisposition and improves 10-year MM risk discrimination from a C-index of 0.68 to 0.74 when incorporated into clinical risk models. These findings invite a conceptual reframing: CRAB-like laboratory deviations are not merely consequences of established malignancy but early, graded signatures of host susceptibility that are detectable in the general population using tests already performed in routine care.
% 我们的发现确立了多发性骨髓瘤的临床前生物学并非从健康到疾病的二元转变，而是一个缓慢演进的、多系统的过程，在临床发病前逾十年便已铭刻于常规实验室检测之中。通过系统描述近30万名前瞻性随访个体的61种血液指标，我们表明蛋白代谢失调、红细胞生成抑制、红细胞形态异质性以及协同性免疫细胞组成转变共同界定了一条可量化的易感性连续谱，该连续谱独立于遗传易感性，纳入临床风险模型后可将MM 10年风险判别从C指数0.68提升至0.74。这些发现引发了一种概念性重构：CRAB样实验室偏离不仅仅是已确立恶性肿瘤的后果，而是可在普通人群中通过常规检查检测到的宿主易感性的早期、分级特征。

The primacy of protein metabolism markers translates an established biological sequence into a population-level, quantifiable signal. Whereas prior prospective serological studies demonstrated that clonal immunoglobulins are detectable up to a decade before MM diagnosis~\cite{Landgren2009MGUS,Weiss2009MGUS}, those studies required specialist assays and were conducted in selected or referred populations. Our data demonstrate for the first time that the same temporal sequence leaves a detectable imprint in two of the most commonly measured routine analytes --- total serum protein and albumin --- in an unselected general-population cohort. Rising total protein and a falling A/G ratio capture the dual hallmark of clonal plasma cell expansion: progressive monoclonal immunoglobulin accumulation alongside suppression of normal polyclonal synthesis. The A/G ratio has previously been validated as an independent prognostic factor for survival in established MM~\cite{Cai2021AGR,Kanapuru2022Disparities}; our results extend its clinical relevance to the pre-diagnostic window in the general population. Critically, dose--response modelling revealed that clinically meaningful MM risk elevation is concentrated in individuals with the most extreme protein values --- a nonlinear threshold feature that could directly inform the design of biomarker-guided referral algorithms, whereby only those with the most pronounced deviations are prioritised for specialist evaluation. The E-values for total protein and A/G ratio (2.46 and 2.63, respectively) substantially exceed what would be biologically plausible as residual confounding, supporting a causal interpretation of these associations.
% 蛋白代谢标志物的主导地位将一个已确立的生物学序列转化为人群层面的可量化信号。尽管此前的前瞻性血清学研究表明克隆性免疫球蛋白在MM诊断前最长达十年即可检测~\cite{Landgren2009MGUS,Weiss2009MGUS}，但这些研究需要专科检测，且在经过选择或转诊的人群中开展。我们的数据首次表明，同一时间序列在两种最常测量的常规指标——血清总蛋白和白蛋白——中留下了可检测的印记，研究对象为未经选择的普通人群队列。总蛋白升高和A/G比值下降捕获了克隆性浆细胞扩增的双重特征：单克隆免疫球蛋白的进行性积累，以及对正常多克隆合成的抑制。A/G比值此前已被验证为已确立MM患者生存的独立预后因素~\cite{Cai2021AGR,Kanapuru2022Disparities}；我们的结果将其临床相关性延伸至普通人群的诊断前窗口期。关键的是，剂量-反应模型揭示，具有临床意义的MM风险升高集中在蛋白值最为极端的个体中——这一非线性阈值特征可直接指导基于生物标志物的转诊算法设计，仅将偏离最为显著的个体优先转诊至专科评估。总蛋白和A/G比值的E值（分别为2.46和2.63）远超残余混杂在生物学上合理的范围，支持对这些关联的因果解读。

Erythropoietic suppression and morphological red cell dysregulation constitute a second independent axis of preclinical MM biology. Clonal plasma cells disrupt normal erythropoiesis through multiple mechanisms: physical displacement of erythroblastic islands, cytokine-mediated induction of erythroblast apoptosis via Fas ligand and TRAIL, and CCL3-driven downregulation of the erythroid master regulators GATA1 and KLF1~\cite{Bouchnita2016Erythropoiesis,Zheng2020CCL3}. The inverse associations of RBC count, haemoglobin, and haematocrit with future MM risk --- each progressive across the full pre-diagnostic decade --- are consistent with this cytokine-mediated marrow displacement beginning well before clinical anaemia is apparent. Concurrently, elevated MCV, RDW, and mean reticulocyte volume likely reflect compensatory erythropoiesis by a shrinking pool of residual erythroid progenitors producing morphologically heterogeneous cells under conditions of marrow infiltration and cytokine stress. The stronger inverse association of RBC count with MM risk in older participants suggests that erythropoietic suppression may be a more sensitive early signal when baseline hematopoietic reserve is already reduced, with potential implications for age-stratified risk thresholds in future clinical applications. Importantly, all of these erythropoietic and morphological indices are components of the standard complete blood count, requiring no additional testing beyond what is already performed in routine clinical care --- a feature that substantially lowers the barrier to implementing erythropoietic signals as part of a population-level MM risk score.
% 红细胞生成抑制和红细胞形态失调构成了临床前MM生物学的第二条独立轴线。克隆性浆细胞通过多种机制扰乱正常红细胞生成：物理性排挤红系造血岛、经Fas配体和TRAIL介导的细胞因子诱导红系前体细胞凋亡，以及CCL3驱动的红系主调控因子GATA1和KLF1下调~\cite{Bouchnita2016Erythropoiesis,Zheng2020CCL3}。红细胞计数、血红蛋白和红细胞压积与未来MM风险的负相关——在整个临床前十年中均呈进行性——与这种细胞因子介导的骨髓替代在临床贫血出现前很久即已开始的推断一致。与此同时，MCV、RDW和平均网织红细胞体积的升高可能反映了在骨髓浸润和细胞因子应激条件下，不断缩小的残余红系祖细胞池通过代偿性造血产生形态异质性红细胞。红细胞计数与MM风险的负相关在年龄较大参与者中更强，提示当基线造血储备已经下降时，红细胞生成抑制可能是更敏感的早期信号，对未来临床应用中年龄分层风险阈值具有潜在意义。重要的是，所有这些红细胞生成和形态学指标均是标准全血细胞计数的组成部分，无需超出常规临床护理范围的额外检测——这一特征大幅降低了将红细胞生成信号纳入人群层面MM风险评分的实施门槛。

A less recognised but informative finding is the coherent shift in circulating immune cell composition that preceded MM diagnosis. Declining neutrophil and monocyte counts alongside rising lymphocyte fractions --- each displaying significant nonlinear dose--response relationships --- are consistent with progressive marrow myeloid suppression and cytokine-mediated immune redistribution driven by the evolving tumour microenvironment. High-dimensional immune profiling studies have shown that immune dysregulation begins as early as the MGUS stage, with alterations in monocyte and dendritic cell function, NK and T cell exhaustion, and expansion of myeloid-derived suppressor cells evident even before clinical progression~\cite{Moscvin2023Immune,Zavidij2020scRNA,Dhodapkar2022Immune}. The decrease in granulocytes in the tumour microenvironment has been specifically identified as predictive of MM outcomes in multi-omics analyses spanning MGUS, SMM, and active MM~\cite{BCJ2024MultiOmics}; our data provide the first evidence that a corresponding shift in peripheral blood granulocyte and monocyte fractions is detectable more than a decade before diagnosis in an unselected general population. SHAP analysis revealed that WBC and neutrophil count contributed mean absolute importances comparable to albumin, underscoring that the immune compositional shift carries additive predictive information beyond dysproteinaemia alone. While definitive mechanistic resolution awaits experimental studies, the temporal precedence of immune shifts over clinically apparent anaemia, and their concordance with single-cell profiling data from the MGUS-to-MM continuum, favour a model in which tumour microenvironment remodelling drives peripheral immune redistribution before marrow infiltration becomes hematologically detectable --- with active immune evasion by the expanding clone and reduced myeloid progenitor output likely acting in concert.
% 一个较少被认识但具有重要信息价值的发现是，先于MM诊断出现的循环免疫细胞组成的协同性转变。中性粒细胞和单核细胞计数下降，同时淋巴细胞比例上升——每项均显示出显著的非线性剂量-反应关系——与不断演变的肿瘤微环境驱动的进行性骨髓髓系抑制和细胞因子介导的免疫重分布一致。高维免疫谱分析研究表明，免疫失调早在MGUS阶段即已开始，单核细胞和树突状细胞功能改变、NK细胞和T细胞耗竭以及髓系来源抑制细胞扩增在临床进展前即已显现~\cite{Moscvin2023Immune,Zavidij2022scRNA,Dhodapkar2022Immune}。肿瘤微环境中粒细胞的减少已在跨越MGUS、SMM和活动性MM的多组学分析中被特别确认为MM预后的预测因素~\cite{BCJ2024MultiOmics}；我们的数据首次提供了在未经选择的普通人群中，相应的外周血粒细胞和单核细胞比例转变在诊断前逾十年即可检测的证据。SHAP分析显示，白细胞和中性粒细胞计数贡献了与白蛋白相当的平均绝对重要性，强调了免疫细胞组成转变在蛋白质代谢异常之外具有附加的预测信息。尽管明确的机制解析有待实验研究，但免疫转变先于临床明显贫血的时间优先性，以及其与MGUS至MM连续谱单细胞谱分析数据的一致性，支持这样一种模型：肿瘤微环境重塑在骨髓浸润变得血液学可检测之前驱动外周免疫重分布——扩增克隆的主动免疫逃逸和骨髓祖细胞输出减少可能协同作用。

The risk prediction analyses provide a practical framework for translating these biomarker findings into clinical action. Protein metabolism markers contributed the largest incremental discrimination gain, consistent with the primacy of paraprotein-driven biology in myelomagenesis. Among the high-risk decile of the biomarker-augmented model, the 10-year cumulative MM incidence reached 0.79\%, compared with 0.47\% in the clinical model alone. To contextualise this absolute risk: the annual MM progression rate from established MGUS is approximately 1\% per year~\cite{Kyle2002MGUS}, and population-level cancer screening programmes for other malignancies have been initiated at comparable or lower absolute risk thresholds. The concentration of risk in a small, identifiable subgroup thus supports a biomarker-guided approach to intensified surveillance, in which individuals with extreme protein and hematological deviations are prioritised for serum protein electrophoresis and free light chain testing~\cite{Smith2022EarlyDiagnosis}. All constituent biomarkers are already measured in routine clinical practice, making the incremental cost of generating a risk score negligible and integration into electronic health record systems technically feasible, consistent with the vision of biomarker-informed cancer surveillance articulated for hematological malignancies more broadly~\cite{Wang2024Screening}.
% 风险预测分析为将这些生物标志物发现转化为临床行动提供了实践框架。蛋白代谢标志物贡献了最大的增量判别提升，与骨髓瘤发生过程中副蛋白驱动生物学的主导地位一致。在生物标志物增强模型的高风险十分位中，10年累积MM发病率达到0.79\%，而单纯临床模型为0.47\%。为将这一绝对风险置于语境中：已确立MGUS的年MM进展率约为1\%/年~\cite{Kyle2002MGUS}，其他恶性肿瘤的人群层面癌症筛查项目已在相当或更低的绝对风险阈值下启动。因此，风险集中在一个小而可识别的亚组中，支持基于生物标志物的强化监测方法，将蛋白质和血液学偏离最为极端的个体优先转诊进行血清蛋白电泳和游离轻链检测~\cite{Smith2022EarlyDiagnosis}。所有组成生物标志物已在常规临床实践中测量，使得生成风险评分的边际成本可忽略不计，并在技术上可行地整合至电子健康记录系统，与更广泛地为血液系统恶性肿瘤所阐明的生物标志物知情癌症监测愿景一致~\cite{Wang2024Screening}。

Our findings must be interpreted in the context of several limitations. The UK Biobank exhibits a well-recognised healthy-volunteer bias and substantially underrepresents non-White ethnic groups, who comprise only approximately 5\% of the analytical cohort. This is a particularly important limitation for MM, in which individuals of African ancestry face approximately twice the age-adjusted incidence of European-ancestry populations, are diagnosed at a younger median age, and carry a disproportionately higher mortality burden~\cite{Bhutani2023Disparities,Duma2019RacialDisparities}. Whether the biomarker signatures identified here operate with the same magnitude and clinical thresholds in populations of African ancestry --- who show distinct MGUS biology including lower M-protein levels and higher prevalence of IgA gammopathy --- cannot be determined from the present data and represents an urgent priority for external validation. Second, biomarkers were measured at a single baseline visit, precluding evaluation of within-individual longitudinal change and the additive value of repeated measurements; the observed preclinical trajectories suggest that dynamic biomarker monitoring over time could substantially improve predictive accuracy beyond cross-sectional ascertainment. Third, with 768 incident MM events, subgroup analyses --- particularly those stratified by ethnicity or within narrow age and sex strata --- are subject to limited statistical power, and the sex-specific heterogeneities identified (stronger albumin and corrected calcium associations in men; $P_{\text{interaction}}=0.004$ for A/G ratio) should be regarded as hypothesis-generating pending validation in larger, more ethnically diverse cohorts. Finally, while the prediction models demonstrated consistent internal performance across repeated cross-validation, external replication in independent prospective cohorts is required before biomarker-guided MM surveillance can be recommended at the population level.
% 我们的发现必须在若干局限性的背景下加以解读。英国生物样本库存在公认的健康志愿者偏倚，且非白种族裔群体严重代表不足，仅占分析队列的约5\%。这对MM而言是一个特别重要的局限性，非洲裔个体面临约为欧洲裔人群两倍的年龄调整发病率，诊断时中位年龄更小，且承受着不成比例的更高死亡负担~\cite{Bhutani2023Disparities,Duma2019RacialDisparities}。本研究数据无法确定此处识别的生物标志物特征是否在非洲裔人群中以相同量级和临床阈值运作——该人群显示出独特的MGUS生物学，包括较低的M蛋白水平和较高的IgA丙种球蛋白病患病率——这代表了外部验证的迫切优先事项。其次，生物标志物在单次基线访视时测量，无法评估个体内纵向变化及重复测量的附加价值；观察到的临床前轨迹提示，随时间进行的动态生物标志物监测可能在横截面确定的基础上显著改善预测准确性。第三，768例新发MM事件使亚组分析——特别是按种族分层或在狭窄的年龄和性别层内的分析——受到有限统计功效的制约，所识别的性别特异性异质性（白蛋白和校正钙与男性的关联更强；A/G比值的$P_{\text{interaction}}=0.004$）应被视为假设生成性发现，有待在更大规模、种族更多样化的队列中验证。最后，尽管预测模型在重复交叉验证中表现出一致的内部性能，在人群层面推荐基于生物标志物的MM监测之前，仍需在独立前瞻性队列中进行外部复制。

In summary, subtle yet systematic deviations in common laboratory tests --- spanning protein metabolism, erythropoiesis, red cell morphology, and immune cell composition --- collectively define a preclinical MM susceptibility continuum that is quantifiable in the general population, independent of inherited genetic predisposition, and detectable more than a decade before clinical diagnosis. These findings support a conceptual reframing of CRAB-like laboratory abnormalities as early, graded signatures of host susceptibility rather than consequences of established disease --- providing a practical foundation for biomarker-guided risk stratification and targeted MM surveillance using tests already embedded in routine clinical care.
% 总之，常见实验室检查中微妙但系统性的偏离——涵盖蛋白代谢、红细胞生成、红细胞形态和免疫细胞组成——共同界定了一条在普通人群中可量化、独立于遗传易感性、且在临床诊断前逾十年即可检测的临床前MM易感性连续谱。这些发现支持对CRAB样实验室异常的概念性重构，将其视为宿主易感性的早期、分级特征，而非已确立疾病的后果——为利用已嵌入常规临床护理的检查进行生物标志物引导的风险分层和靶向MM监测提供了实践基础。

\section*{Methods}

\subsection*{Study population}

This study utilized data from the UK Biobank, a large-scale prospective cohort that recruited approximately 500,000 participants aged 40–69 years across 22 assessment centres in England, Scotland, and Wales between 2006 and 2010. All participants provided written informed consent at enrolment, and the study received ethical approval from the North West Multi-Centre Research Ethics Committee (reference 11/NW/0382). Baseline data collection included sociodemographic information, lifestyle questionnaires, physical measurements, and biological samples including blood, urine, and saliva.

For the present analysis, we excluded participants with a diagnosis of multiple myeloma (MM) at or before the baseline assessment date, those with any cancer diagnosis prior to baseline, and those with missing or invalid follow-up information (follow-up time $\leq$ 0 or undetermined event status). The final analytical cohort was further restricted to participants with complete data across all retained blood biomarkers (biomarker complete-case; see below). All analyses were conducted in accordance with UK Biobank data access protocol (application number 146760).

\subsection*{Blood biomarker measurements}

Venous blood samples were collected at the baseline assessment visit using standardized protocols and processed at a central laboratory. The UK Biobank measured a comprehensive panel of hematological and biochemical markers. For this study, we considered up to 61 candidate blood biomarkers spanning six physiological domains: (1) \textit{Inflammation} (white blood cell count, neutrophil count, neutrophil percentage, lymphocyte count, lymphocyte percentage, monocyte count, monocyte percentage, eosinophil count, eosinophil percentage, basophil count, basophil percentage, C-reactive protein, nucleated red blood cell count and percentage); (2) \textit{Hematology — Count} (red blood cell count, haemoglobin, haematocrit, platelet count, reticulocyte count, reticulocyte percentage, high-scatter reticulocyte count and percentage, immature reticulocyte fraction); (3) \textit{Hematology — Morphology} (mean corpuscular volume, mean corpuscular haemoglobin, mean corpuscular haemoglobin concentration, red blood cell width, mean reticulocyte volume, mean sphered cell volume, platelet crit, platelet distribution width, mean platelet volume); (4) \textit{Protein Metabolism} (total protein, albumin); (5) \textit{Kidney} (cystatin C, creatinine, urea, urate); (6) \textit{Hepatobiliary} (direct bilirubin, total bilirubin, alanine aminotransferase, aspartate aminotransferase, alkaline phosphatase, gamma-glutamyltransferase); and (7) \textit{Lipid/Nutrition} (cholesterol, LDL-cholesterol, HDL-cholesterol, triglycerides, apolipoprotein A, apolipoprotein B, lipoprotein(a), glucose, glycated haemoglobin); along with additional biomarkers including calcium, phosphate, vitamin D, testosterone, oestradiol, SHBG, IGF-1, and rheumatoid factor.

Two derived biomarkers were computed prior to any sample restriction: the albumin-to-globulin ratio (albumin / [total protein $-$ albumin], defined only when total protein $>$ albumin) and corrected calcium (calcium + 0.02 $\times$ [40 $-$ albumin], using the Payne correction formula. The units for corrected calcium and albumin are mmol/L and g/L, respectively). Biomarkers with more than 20\% missing values in the post-exclusion cohort were removed from all analyses (missing threshold: 20\%). This criterion was applied identically across all analytical modules to ensure consistency. Biomarkers passing this threshold (denoted as ``retained biomarkers'') were recorded and used uniformly throughout the study. For the complete-case biomarker dataset used in the main analysis, only participants with non-missing values for all retained biomarkers simultaneously were included.

\subsection*{Ascertainment of multiple myeloma}

Incident MM cases were identified through linkage to national cancer registries (NHS Digital for England and Wales; Information Services Division for Scotland) and hospital episode statistics using International Classification of Diseases, 10th revision (ICD-10) code C90.0. The date of first MM diagnosis was extracted from the earliest available record across linked data sources. Participants were classified as incident MM cases if their diagnosis occurred strictly after the baseline assessment date and within the observation period (through 1 January 2024). Participants with a diagnosis date on or before baseline were excluded as prevalent cases.

\subsection*{Follow-up and event definition}

Follow-up time was calculated from the baseline assessment date to the earliest of: date of MM diagnosis, date of death (from the UK Biobank death register), date of loss to follow-up (withdrawal from the study), or the administrative censoring date of 1 January 2024. Participants who died before MM diagnosis were censored at their date of death. A binary event indicator was defined as 1 if incident MM was confirmed within the follow-up window, and 0 otherwise. Participants with non-positive follow-up time or indeterminate event status were excluded.

\subsection*{Covariate definitions}

The following covariates were considered across all regression models, organised into two nested model tiers:

\textbf{Model 1 (sociodemographic + lifestyle):} age at baseline (continuous, years), sex (male/female), self-reported ethnicity (White/Other), Townsend deprivation index (TDI; continuous), educational attainment (college/university degree: yes/no), and annual household income (categorical), smoking status (never/previous/current), alcohol consumption frequency (categorical), physical activity (International Physical Activity Questionnaire [IPAQ] metabolic equivalent of task [MET] categories: low/middle/high), and sleep duration (categorical).

\textbf{Model 2 (Model 1 + health status):} additionally included body mass index (BMI; kg/m$^2$), prevalent cardiovascular disease at baseline (history of ICD-10 codes I20–I25, I50, I60–I64: yes/no), prevalent type 2 diabetes (ICD-10 E11: yes/no), prevalent hypertension (ICD-10 I10–I13, I15: yes/no), family history of cancer (yes/no), and polygenic risk score for MM (PRS; standardized sum score, continuously; and PRS tertile: T1-Low/T2-Medium/T3-High). Comorbidity dates were ascertained through linked hospital episode statistics, and prevalent status was defined as any diagnosis prior to the baseline date. The polygenic risk score was derived from genome-wide association study summary statistics and standardized to zero mean and unit variance within the analytical cohort.

\textbf{Genetic Risk Assessment}: 
We employed a clustering and thresholding (C+T) approach to conduct polygenic risk score (PRS) analysis on imputed data from the UK Biobank (UKB) genome-wide association study (GWAS). First, we subjected the UKB imputed data to stringent quality control: 
(i) SNP removal threshold of $0.01$, 
(ii) individual removal threshold of $0.02$, 
(iii) minor allele frequency (MAF) threshold of $0.01$, and 
(iv) Hardy-Weinberg equilibrium threshold of $1 \times 10^{-6}$. Subsequently, we systematically searched GWAS catalogues and extracted genome-wide significant variants for myeloma from the largest GWAS study (\texttt{finngen\_R12\_C3\_MULT\_MYELOMA\_EXALLC}). Using the C+T approach (PLINK 2.0), we generated PRS for multiple thresholds ($P < 5 \times 10^{-8}$, $5 \times 10^{-6}$, $1 \times 10^{-5}$) for the myeloma trait to determine the optimal genetic risk assessment strategy. For the association analysis, the calculated PRS was standardized and then categorized into tertiles (low, intermediate, and high genetic risk groups). We evaluated the risk of myeloma by comparing the high and intermediate risk groups against the lowest tertile (reference group) using regression models. The principles of PRS calculation in PLINK 2.0 and the genotyping and imputation procedures for SNPs in the UK Biobank are detailed in \cite{wang2024associations}.

\textbf{Handling of missing covariates:} Missingness in covariates was assessed after biomarker complete-case restriction. Continuous variables with $\leq$5\% missing data were imputed with the within-sample median; categorical variables with $\leq$5\% missing were imputed with the within-sample mode. For variables with $>$5\% missingness, the same imputation strategy was applied with the additional creation of a binary missingness indicator variable (``Missing'' / ``Not Missing'') that was included as an additional covariate in all models. The PRS continuous score was imputed to the median if missing, after which PRS tertile groups were recomputed using the \texttt{ntile()} function on the imputed values.

\subsection*{Baseline characteristics}

Baseline characteristics of the analytical cohort were summarised overall and stratified by MM status (incident MM versus no MM) using the \texttt{tableone} package in R. Continuous variables are reported as median (interquartile range) or mean (standard deviation) as appropriate; categorical variables are reported as number (percentage). Group comparisons employed the standardized mean difference for continuous variables and chi-squared or Fisher's exact tests for categorical variables.

\subsection*{Biomarker-wide association scan (PheWAS-style analysis)}

To systematically evaluate associations between circulating blood biomarkers and incident MM risk, we conducted a phenome-wide association study (PheWAS)-style biomarker scan. For each of the retained biomarkers, we fitted three nested multivariable Cox proportional hazards regression models (Models 1–3 as defined above) using the \texttt{survival} package in R. Each biomarker was standardized to a mean of zero and unit standard deviation (Z-score) within the analytical sample prior to entry into the model, so that hazard ratios (HRs) represent the change in MM risk per one standard deviation (SD) increase in the biomarker. The resulting HR, 95\% confidence interval (CI), and two-sided Wald $p$-value were extracted for each biomarker–model combination. Analyses were restricted to biomarker-complete-case records (non-missing for the specific biomarker being tested), with the requirement of $\geq$1,000 participants and $\geq$10 incident MM events per model fit.

\textbf{Multiple testing correction:} Given the large number of biomarkers tested, $p$-values from each model tier were corrected for multiple comparisons using the Benjamini–Hochberg false discovery rate (FDR) procedure, applied separately within each model across all biomarkers. Biomarkers with FDR-corrected $p < 0.05$ were considered statistically significant. Bonferroni-corrected thresholds are additionally reported. The $-\log_{10}$(FDR) values for biomarkers with extreme significance were plotted using a compressed $y$-axis transformation to enhance readability: $y_{\text{plot}} = y$ for $y \leq 6$, and $y_{\text{plot}} = 6 + (y - 6) \times 0.18$ for $y > 6$.

\subsection*{Targeted biomarker Cox regression}

In addition to the full biomarker-wide scan, we conducted pre-specified targeted analyses for 20 biomarkers representing six pathophysiologically relevant categories for MM: \textit{Inflammation} (white blood cell count, neutrophil count, neutrophil percentage, lymphocyte percentage, monocyte count, eosinophil count), \textit{Hematology — Count} (red blood cell count, haemoglobin, haematocrit, reticulocyte count), \textit{Hematology — Morphology} (mean reticulocyte volume, mean corpuscular volume, mean corpuscular haemoglobin, red blood cell width, mean sphered cell volume), \textit{Protein Metabolism} (total protein, albumin, albumin-to-globulin ratio), \textit{Kidney} (cystatin C), and \textit{Others} (corrected calcium). These targeted biomarkers were selected a priori based on their biological relevance to plasma cell biology and MM pathogenesis. For each biomarker, Models 1–2 were fitted as described above, and FDR correction was applied within each model tier across all targeted biomarkers.

\subsection*{Stratified analyses and tests for interaction}

To evaluate potential effect modification, we conducted stratified Cox regression analyses for the pre-specified targeted biomarkers across four subgroup variables: PRS tertile (T1-Low, T2-Medium, T3-High), sex (male, female), age group ($<$65 years, $\geq$65 years), and BMI group ($<$25 kg/m$^2$, $\geq$25 kg/m$^2$). Within each stratum, Model 2 covariates were fitted with the stratifying variable removed to avoid collinearity (e.g., age was excluded as a continuous covariate in age-group stratified analyses; sex was excluded in sex-stratified analyses). To formally test for interaction, we fitted a multiplicative interaction model of the form:

\begin{equation*}
h(t) = h_0(t) \exp(\beta_1 Z + \beta_2 S_{\text{numeric}} + \beta_3 Z \times S_{\text{numeric}} + \boldsymbol{\gamma}^{\top} \mathbf{X})
\end{equation*}

where $Z$ is the biomarker Z-score, $S_{\text{numeric}}$ is the stratifying variable coded as an ordinal numeric score, and $\mathbf{X}$ is the vector of remaining covariates. The coefficient $\beta_3$ and its corresponding Wald $p$-value were used as the interaction test. FDR correction for interaction $p$-values was applied separately within each stratification analysis. A minimum of 50 participants with at least 3 MM events per stratum was required.

\subsection*{Residual biomarker analysis}

To disentangle the contribution of genetic predisposition (captured by PRS) from environmentally modifiable biomarker variation, we performed a two-stage residual analysis for all targeted biomarkers. In the first stage, we regressed each biomarker on the standardized PRS continuous score using ordinary least squares regression (\texttt{lm} in R). The coefficient of determination ($R^2$) was recorded as a measure of PRS-explained variance in the biomarker. In the second stage, residuals from each regression (i.e., the PRS-independent component of each biomarker) were standardized to Z-scores and entered as the exposure in Cox models. Model 2 covariates were used in the second stage, with PRS deliberately excluded to avoid double-adjustment. FDR correction was applied separately within each analysis stage. This two-stage approach allows assessment of whether biomarker associations with MM risk persist independently of genetic liability to MM.

\subsection*{Dose–response analysis using restricted cubic splines}

To characterise the shape of the association between each targeted biomarker and MM incidence (linear versus non-linear), we fitted restricted cubic spline (RCS) models using the \texttt{rms} package in R. For each biomarker, a Cox model with three knots placed at default quantile positions was fitted:

\begin{equation*}
h(t) = h_0(t) \exp\bigl[\text{rcs}(X, 3) + \boldsymbol{\gamma}^{\top} \mathbf{X}_{\text{cov}}\bigr]
\end{equation*}

where $\text{rcs}(X, 3)$ denotes the RCS basis expansion of the biomarker with 3 knots and $\mathbf{X}_{\text{cov}}$ denotes the Model 2 covariate vector. Models were fitted on both the original scale and the Z-score scale. The overall association $p$-value ($P_{\text{overall}}$) and the $p$-value for departure from linearity ($P_{\text{nonlinear}}$) were extracted from the type III analysis of variance table (\texttt{anova.rms}). A smooth HR curve with 95\% CI was generated by predicting the linear predictor across a grid of 200 evenly spaced values between the 0.5th and 99.5th percentiles of the observed distribution, referenced to the sample median (original scale) or zero (Z-score scale). Kernel density estimates of the biomarker distribution were overlaid on each panel, scaled proportionally to the maximum observed HR. For MM cases, a spike rug plot (truncated to a random sample of 500 if $n > 500$) was added to depict the distribution of biomarker values among those who developed MM. A minimum sample size of 500 was required for an RCS fit to be reported.

\subsection*{Biomarker trajectory analysis}

To examine whether biomarker levels at baseline differed according to the proximity of subsequent MM diagnosis, we categorised incident MM cases into four groups based on time from baseline to diagnosis: 0–3 years, 4–7 years, 8–11 years, and $\geq$11 years. Participants without an MM diagnosis served as the reference group (``No MM''). Participants were excluded from this analysis if their diagnosis date preceded or coincided with the baseline date.

\textbf{Descriptive statistics:} Mean and standard deviation of each biomarker were computed within each trajectory group.

\textbf{Between-group testing:} One-way ANOVA was performed for each biomarker across the four MM trajectory groups (excluding the No MM reference group), with FDR correction applied across all biomarkers.

\textbf{Ordered-group linear regression:} To formally test for a monotonic trend in biomarker levels as a function of time to diagnosis, we fitted a linear regression model of the form: $\text{biomarker} = \beta_0 + \beta_1 G + \beta_2 \text{age} + \beta_3 \text{sex} + \varepsilon$, where $G$ is a numeric group score coded as: No MM = 0, $\geq$11 years = 1, 8–11 years = 2, 4–7 years = 3, 0–3 years = 4. A positive $\beta_1$ therefore indicates higher biomarker values closer to MM diagnosis. FDR correction was applied across all biomarkers.

\subsection*{Prediction model development and evaluation}

We evaluated the incremental value of blood biomarkers for predicting MM onset at 5-year and 10-year horizons beyond established clinical risk factors. Two prognostic Cox models were constructed:

\begin{itemize}
    \item \textbf{clinical model:} age, sex, ethnicity, educational attainment, household income, Townsend deprivation index, smoking status, alcohol frequency, physical activity, sleep duration, prevalent CVD, prevalent type 2 diabetes, prevalent hypertension, family history of cancer, and BMI.
    \item \textbf{clinical + biomarker model:} clinical model variables plus all 20 pre-specified targeted biomarkers.
\end{itemize}

\textbf{Cross-validation procedure:} Model performance was evaluated using a repeated random split-sample approach with 10 iterations. In each iteration, 70\% of participants were randomly assigned to a training set (seed = 123 + iteration index) and the remaining 30\% constituted the test set. Each Cox model was fitted on the training set, and the linear predictor was obtained for the test set. Model discrimination was assessed using: (1) the time-truncated Harrell's concordance index (C-index), computed against the time-to-event outcome censored at the respective prediction horizon (5 or 10 years); and (2) the time-dependent area under the receiver operating characteristic curve (AUC) at the specified horizon, estimated using inverse probability of censoring weighting (IPCW) as implemented in the \texttt{timeROC} package. Sensitivity and specificity at the optimal Youden index threshold were also recorded.

\textbf{Risk stratification:} In the first iteration, test-set participants were classified into Low (bottom 10\%), Intermediate (10th–90th percentile), and High (top 10\%) risk groups based on decile cut-points of the linear predictor. Cumulative MM incidence curves stratified by risk group were constructed using the Kaplan–Meier estimator applied to the time-truncated event indicator, and group differences were assessed using the log-rank test.

\textbf{Performance reporting:} C-indices and AUC values are reported as mean $\pm$ standard deviation across 10 iterations; C-index 95\% CI is computed as mean $\pm$ 1.96 $\times$ standard error (standard deviation / $\sqrt{10}$).

\subsection*{Sensitivity analyses}

Three pre-specified sensitivity analyses were conducted to assess the robustness of the main findings from targeted biomarker Cox analyses:

\textbf{(1) Two-year landmark exclusion:} Participants who developed MM within 2 years of baseline were excluded from the case group to reduce potential reverse causation bias, whereby subclinical disease at the time of blood sampling might artificially inflate associations.

\textbf{(2) Complete-case analysis:} Rather than using the imputed main analysis dataset, we repeated all Cox analyses on the biomarker complete-case dataset (i.e., participants with non-missing values across all retained biomarkers simultaneously) without the use of missingness indicator covariates.

\textbf{(3) Fine-Gray competing risks model:} To account for the competing risk of death before MM diagnosis, we fitted Fine-Gray subdistribution hazard models using the \texttt{cmprsk} package. Participants who died without MM were assigned a competing event code of 2 (versus 1 for MM, 0 for censored). The biomarker Z-score and Model 2 covariates were entered as covariates via the design matrix (\texttt{model.matrix}). FDR correction within each analysis-model stratum was applied consistently.

\textbf{E-value analysis:} To quantify the robustness of significant findings to potential unmeasured confounding, we computed E-values for all biomarker associations that remained statistically significant (FDR $< 0.05$) across any sensitivity analysis. The E-value is defined as the minimum strength of association (on the risk ratio scale) that an unmeasured confounder would need to have with both the biomarker exposure and MM outcome — above and beyond the measured covariates — to fully explain away the observed association~\cite{VanderWeele2017}. For hazard ratios approximated as risk ratios, the E-value was computed as:

\begin{equation*}
\text{E-value} = \text{RR} + \sqrt{\text{RR} \times (\text{RR} - 1)}
\end{equation*}

where RR = HR for associations with HR $\geq$ 1, and RR = 1/HR for protective associations (HR $< 1$). An E-value was also computed for the confidence interval limit closest to the null (lower CI for HR $> 1$; upper CI for HR $< 1$) to indicate the minimum confounding strength needed to shift the confidence interval to include the null.

\subsection*{Statistical software}

All analyses were performed in R (version 4.5; R Foundation for Statistical Computing, Vienna, Austria). Key packages included \texttt{survival} (Cox regression), \texttt{rms} (restricted cubic splines), \texttt{cmprsk} (Fine-Gray models), \texttt{timeROC} (time-dependent AUC), \texttt{tableone} (baseline characteristics), \texttt{forestploter} (forest plots), \texttt{ggplot2}, \texttt{patchwork}, \texttt{cowplot} (visualisation), and \texttt{ggrepel} (label placement). All statistical tests were two-sided; $p < 0.05$ was used as the threshold for nominal significance. Multiple testing was corrected using the Benjamini–Hochberg FDR procedure unless otherwise stated. No imputation was applied to blood biomarker values; all biomarker analyses used complete-case participants for the biomarker(s) under study.

\section*{Data availability}
The UK Biobank data used in this study were accessed under application ID 146760. 
Researchers may apply for access via the UK Biobank website (https://www.ukbiobank.ac.uk/). 
Derived data supporting the findings of this study are available from the corresponding author upon reasonable request, in accordance with UK Biobank data access policies.

\section*{Code availability}
The code used to generate the results in this study is available from the corresponding author upon reasonable request. 
An interactive web-based application for risk prediction is publicly available at \href{https://myelomarisk.org}{myelomarisk.org}.

\section*{Acknowledgements}
The authors acknowledge support from the Singapore Ministry of Education Tier 2 grant A-8001562-00-00 and the Tier 1 grants A-8002931-00-00 and A-8004146-00-00 at the National University of Singapore, as well as the Interdisciplinary Collaborative Research Project of the Shanghai Institute for Mathematics and Interdisciplinary Sciences (SIMIS) (Grant No. SIMIS-ID-2025-OM).

% \section*{Author contributions}
 \section*{Competing interests}
 The authors declare no competing interests.

\bibliographystyle{unsrt}
\bibliography{references} % replace with your .bib file

\clearpage
\section*{Tables and Figures}

\clearpage
\begin{table}[t]
  \centering
  \caption{Baseline characteristics of UK Biobank participants by multiple myeloma status.}
  \label{table1}
  \vspace{5pt}
  \renewcommand{\arraystretch}{1}
  \resizebox{\textwidth}{!}{
  \begin{tabular}{lllll}
  \toprule
  \textbf{Characteristic} & \textbf{Level} & \textbf{No MM (n=298,267)} & \textbf{Incident MM (n=768)} & \textbf{P-value} \\
  \midrule

  \textbf{Sociodemographic characteristics} & & & & \\

  Age, years & Mean (SD) & 55.98 (8.17) & 60.58 (6.81) & <0.001 \\

  Sex, n (\%) & Female & 142{,}294 (47.7) & 269 (35.0) & \multirow{2}{*}{<0.001} \\
              & Male   & 155{,}973 (52.3) & 499 (65.0) & \\

  Ethnicity, n (\%) & Non-white & 15{,}167 (5.1) & 35 (4.6) & 0.560 \\
                    & White     & 283{,}100 (94.9) & 733 (95.4) & \\

  College education, n (\%) & No  & 200{,}144 (67.1) & 512 (66.7) & 0.827 \\
                            & Yes & 98{,}123 (32.9) & 256 (33.3) & \\

  Household income, n (\%) 
      & <£18,000 & 55{,}086 (18.5) & 176 (22.9) & \multirow{5}{*}{<0.001} \\
      & £18,000–30,999 & 63{,}729 (21.4) & 177 (23.0) & \\
      & £31,000–51,999 & 109{,}541 (36.7) & 292 (38.0) & \\
      & £52,000–100,000 & 55{,}188 (18.5) & 100 (13.0) & \\
      & >£100,000 & 14{,}723 (4.9) & 23 (3.0) & \\

  Townsend Deprivation Index & Median [IQR] & $-2.16 [-3.65, 0.50]$ & $-2.38 [-3.81, 0.02]$ & 0.011 \\

  \textbf{Lifestyle factors} & & & & \\

  Smoking status, n (\%) & Non-smoker & 166{,}278 (55.7) & 413 (53.8) & 0.288 \\
                         & Smoker     & 131{,}989 (44.3) & 355 (46.2) & \\

  Drinking frequency, n (\%) 
      & Daily/almost daily & 61{,}343 (20.6) & 146 (19.0) & \multirow{6}{*}{0.062} \\
      & 3–4 times/week & 70{,}718 (23.7) & 170 (22.1) & \\
      & 1–2 times/week & 78{,}761 (26.4) & 221 (28.8) & \\
      & 1–3 times/month & 32{,}766 (11.0) & 74 (9.6) & \\
      & Special occasions & 32{,}309 (10.8) & 81 (10.5) & \\
      & Never & 22{,}370 (7.5) & 76 (9.9) & \\

  Sleep duration, n (\%) 
      & Long sleep   & 21{,}700 (7.3)  & 75 (9.8) & \multirow{3}{*}{0.029} \\
      & Normal sleep & 201{,}994 (67.7) & 508 (66.1) & \\
      & Short sleep  & 74{,}573 (25.0) & 185 (24.1) & \\

  Physical activity, n (\%) 
      & Low & 43{,}210 (14.5) & 113 (14.7) & \multirow{3}{*}{0.981} \\
      & Middle & 94{,}207 (31.6) & 241 (31.4) & \\
      & High & 160{,}850 (53.9) & 414 (53.9) & \\

  \textbf{Clinical measurements} & & & & \\

  Body mass index, kg/m$^{2}$ & Mean (SD) & 27.53 (4.75) & 27.98 (4.51) & 0.009 \\

  \textbf{Comorbidity history} & & & & \\

  Cardiovascular disease, n (\%) 
      & No  & 277{,}538 (93.1) & 692 (90.1) & 0.002 \\
      & Yes & 20{,}729 (6.9)   & 76 (9.9)   & \\

  Type 2 diabetes, n (\%) 
      & No  & 290{,}293 (97.3) & 749 (97.5) & 0.818 \\
      & Yes & 7{,}974 (2.7)    & 19 (2.5)   & \\

  Hypertension, n (\%) 
      & No  & 218{,}195 (73.2) & 522 (68.0) & 0.001 \\
      & Yes & 80{,}072 (26.8) & 246 (32.0) & \\

  \textbf{Family history and genetic factors} & & & & \\

  Family history of cancer, n (\%) 
      & No  & 195{,}587 (65.6) & 494 (64.3) & 0.490 \\
      & Yes & 102{,}680 (34.4) & 274 (35.7) & \\

  Polygenic risk score tertile, n (\%) 
      & Low (T1) & 99{,}463 (33.3) & 216 (28.1) & \multirow{3}{*}{<0.001} \\
      & Middle (T2) & 99{,}435 (33.3) & 243 (31.6) & \\
      & High (T3) & 99{,}369 (33.3) & 309 (40.2) & \\

  \bottomrule
  \end{tabular}}
  
  \vspace{4pt}
  \raggedright
  \footnotesize
  Data presented as n (\%) for categorical variables, mean (standard deviation) for normally distributed variables, 
  and median [interquartile range] for non-normally distributed variables. 
  P-values calculated using $\chi^2$ test (categorical), t-test (normal continuous), and Mann–Whitney U test (non-normal continuous).
  MM: multiple myeloma; SD: standard deviation; IQR: interquartile range.
\end{table}
\clearpage

\clearpage
    {\centering
    \includegraphics[width=\textwidth]{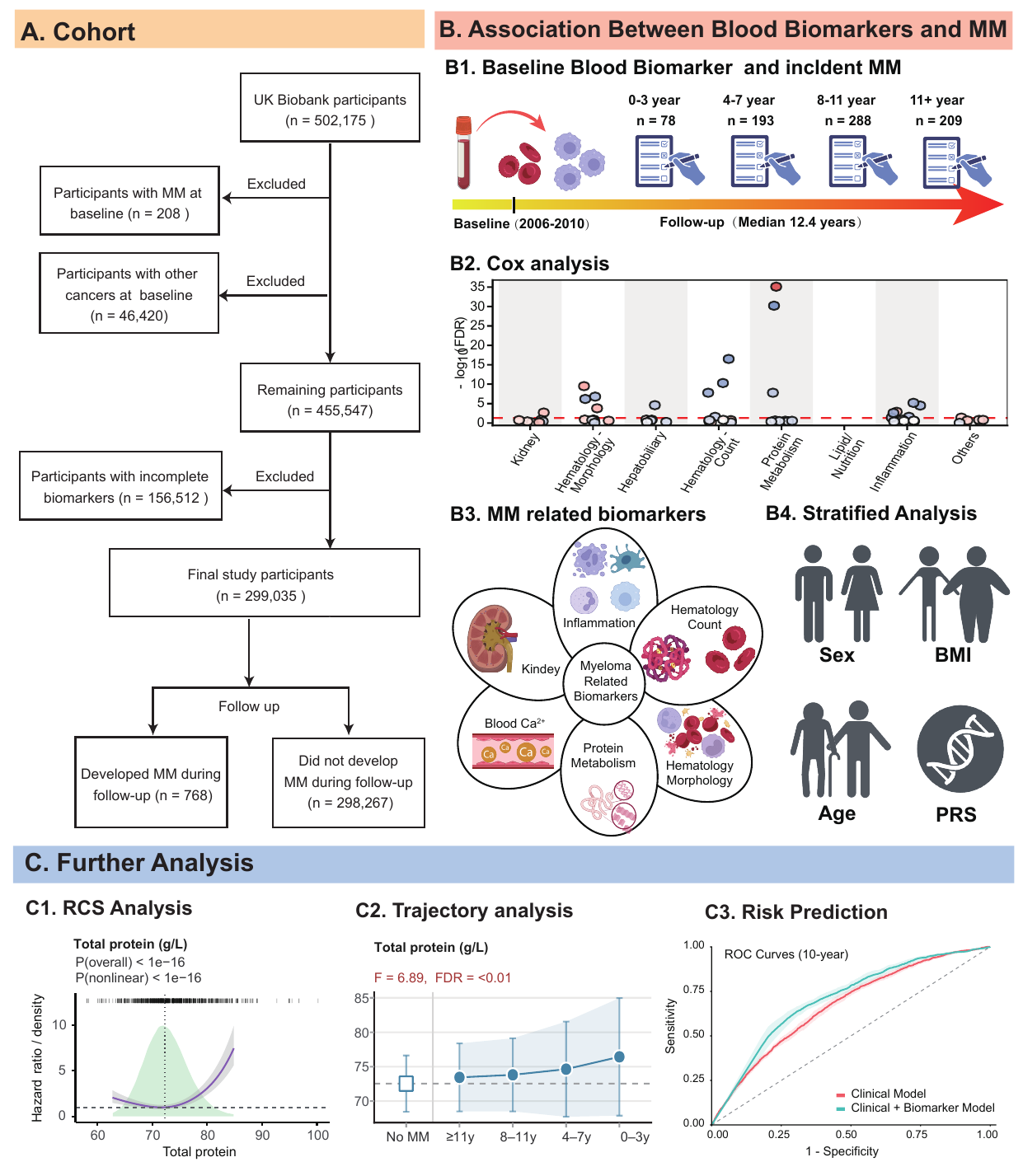}
    \captionof{figure}{\textbf{Study design and overview of associations between blood biomarkers and incident multiple myeloma.}
    \textbf{(A)} Flow diagram of participant selection from the UK Biobank. Of 502,175 enrolled participants, 208 with prevalent multiple myeloma (MM) and 46,420 with other cancers at baseline were excluded, followed by exclusion of 156,512 participants with incomplete biomarker data, yielding a final cohort of 299,035 individuals. During a median follow-up of 12.4 years (2006--2010 to last contact), 768 participants developed incident MM.
    \textbf{(B)} Association between baseline blood biomarkers and incident MM. \textbf{B1}, Timeline illustrating the distribution of MM diagnoses across follow-up intervals (0--3 years, $n=78$; 4--7 years, $n=193$; 8--11 years, $n=288$; $\geq$11 years, $n=209$). \textbf{B2}, Volcano-style plot of Cox proportional hazards analyses showing $-\log_{10}$(FDR-adjusted $P$-values) for 35 blood biomarkers grouped by category (kidney function, hematology morphology, hepatobiliary, hematology count, protein metabolism, lipid/nutrition, inflammation, and others); the dashed red line indicates the FDR significance threshold. \textbf{B3}, Schematic summarizing MM-related biomarker categories identified in Cox analyses, including inflammation, hematology count, hematology morphology, protein metabolism, blood calcium, and kidney function. \textbf{B4}, Stratified analyses were conducted by sex, BMI, age, and polygenic risk score (PRS).
    \textbf{(C)} Further analyses. \textbf{C1}, Restricted cubic spline (RCS) analysis of total protein (g/L) and MM hazard ratio, demonstrating a significant nonlinear dose--response relationship ($P_{\text{overall}}<1\times10^{-16}$; $P_{\text{nonlinear}}<1\times10^{-16}$). \textbf{C2}, Latent class trajectory analysis of total protein levels across four time windows prior to MM diagnosis ($\geq$11 years, 8--11 years, 4--7 years, and 0--3 years before diagnosis) compared with non-MM controls, revealing a progressive elevation in total protein among future MM cases ($F=6.89$; FDR$<0.01$). \textbf{C3}, Receiver operating characteristic (ROC) curves for 10-year MM risk prediction, comparing the clinical model alone (pink) versus the combined clinical and biomarker model (teal); the biomarker-augmented model demonstrated improved discriminative performance.
    MM: multiple myeloma; FDR: false discovery rate; BMI: body mass index; PRS: polygenic risk score; RCS: restricted cubic spline; ROC: receiver operating characteristic.}
    \label{fig:overview}
    }

\clearpage
    {\centering
\includegraphics[width=\textwidth]{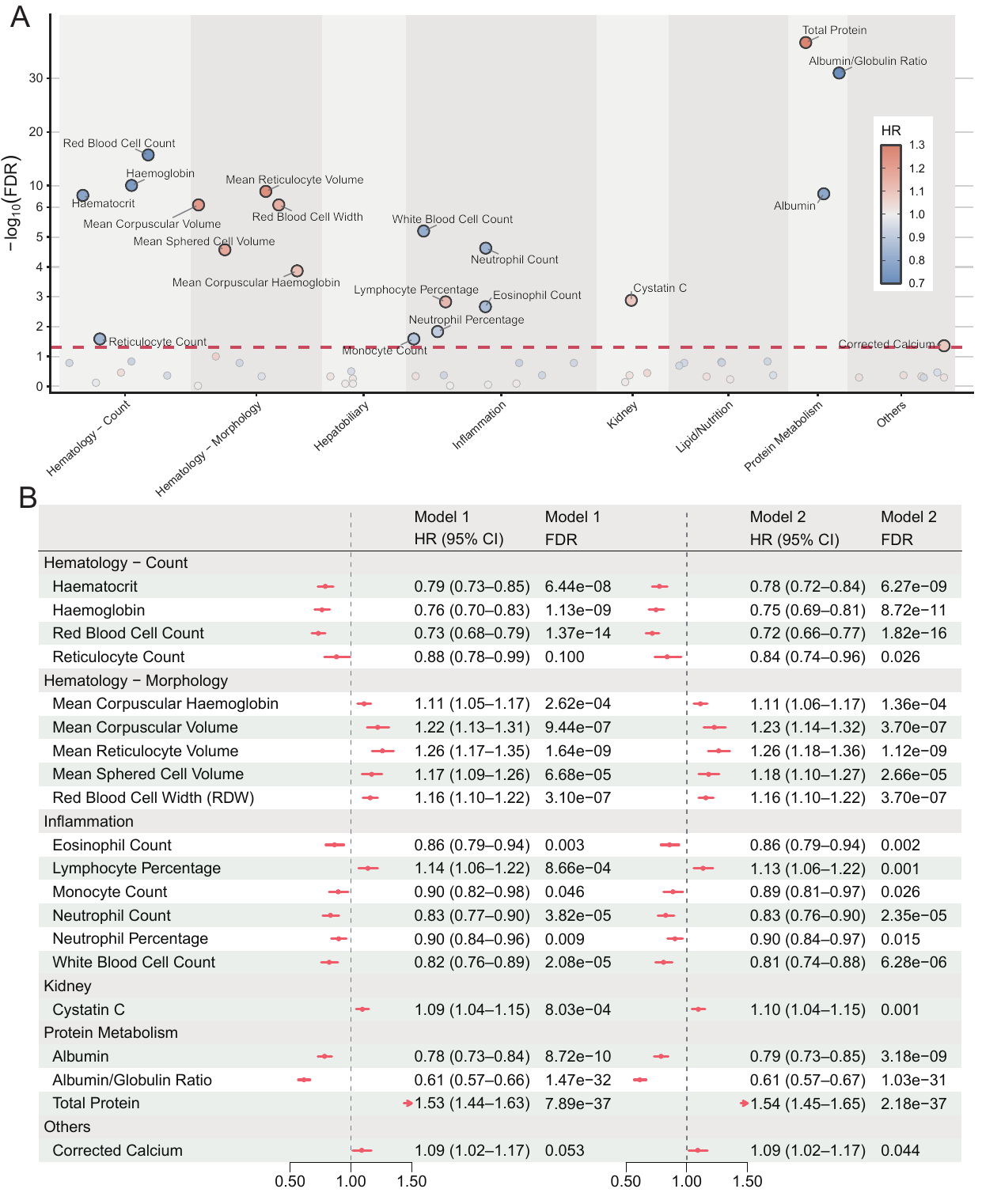}
\captionof{figure}{\textbf{Associations between baseline blood biomarkers and incident multiple myeloma.}
\textbf{(A)} Volcano-style plot of Cox proportional hazards analyses showing $-\log_{10}$(FDR-adjusted $P$-values) on the $y$-axis for 59 blood biomarkers grouped by category (kidney function, hematology morphology, hepatobiliary, hematology count, protein metabolism, lipid/nutrition, inflammation, and others). Dot colour encodes the direction and magnitude of the hazard ratio (HR); the dashed red line indicates the FDR significance threshold ($\alpha = 0.05$). Biomarkers surpassing this threshold are labeled.
\textbf{(B)} Forest plot displaying HR (95\% confidence interval) and FDR-adjusted $P$-values for all FDR-significant biomarkers under Model~1 (adjusted for age, sex, ethnicity, educational attainment, household income, and Townsend Deprivation Index) and Model~2 (Model~1 additionally adjusted for smoking status, alcohol consumption frequency, physical activity, sleep duration, BMI, prevalent cardiovascular disease, type~2 diabetes, hypertension, family history of cancer, and polygenic risk score). All biomarkers were $z$-score standardized; HRs represent the change in MM risk per 1-SD increase in biomarker level.
FDR: false discovery rate; HR: hazard ratio; CI: confidence interval; SD: standard deviation; RDW: red blood cell distribution width.}
\label{fig:biomarker_mm}
}

\clearpage
    {\centering
    \includegraphics[width=\textwidth]{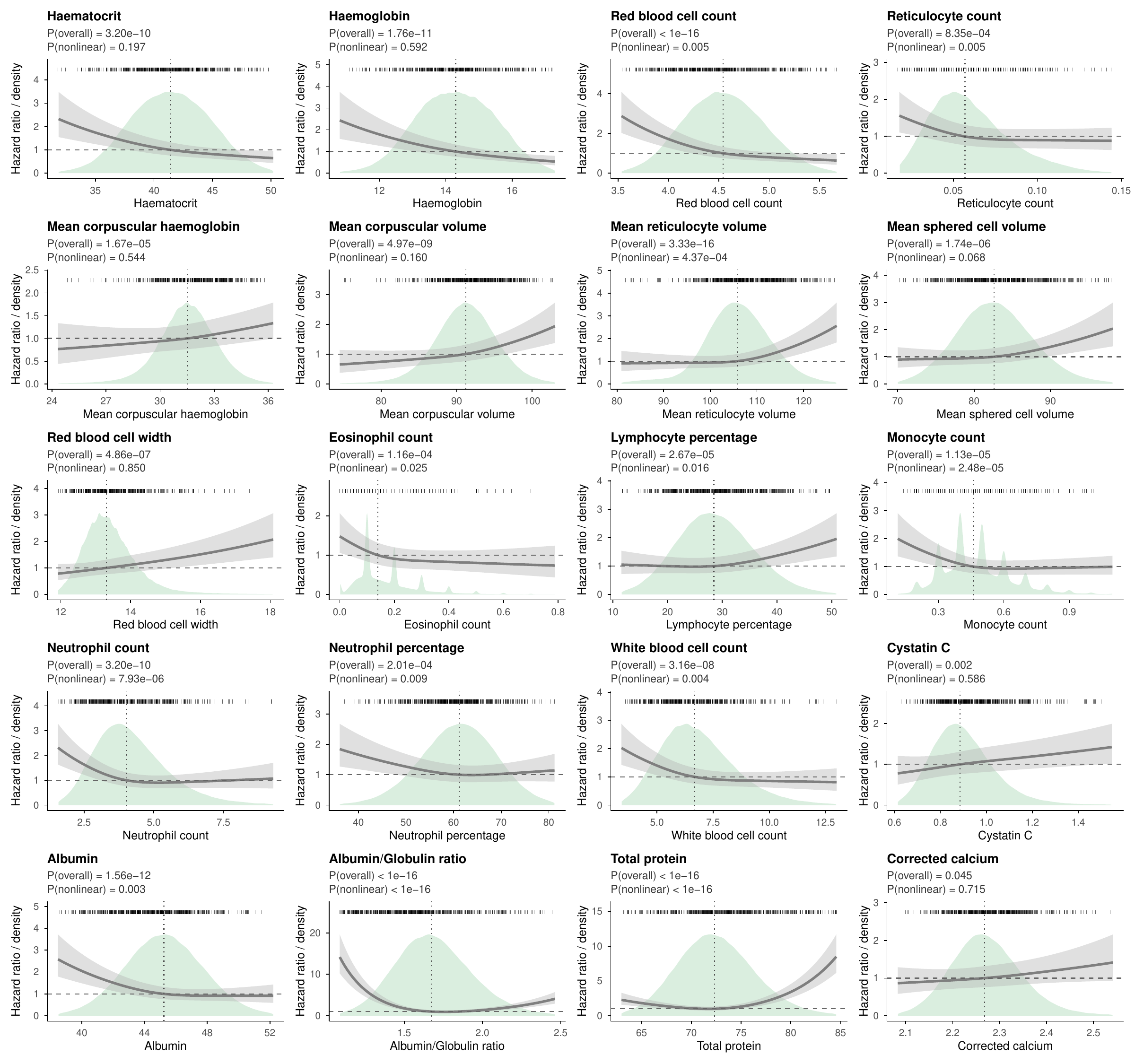}
    \captionof{figure}{\textbf{Nonlinear dose--response associations between baseline biomarker levels and incident multiple myeloma.}
    Restricted cubic spline (RCS) models were used to estimate hazard ratios (HRs) for incident MM across the observed distribution of each biomarker, adjusted for all Model~2 covariates. HRs are referenced to the sample median of each biomarker (vertical dotted line). Shaded grey areas indicate 95\% confidence intervals. Green density curves represent the population distribution of each biomarker. Black tick marks at the top of each panel denote the biomarker values of incident MM cases. Curve colours indicate biomarker category: red, hematology count; green, hematology morphology; blue, inflammation; purple, protein metabolism; orange, kidney; brown, others. $P_{\text{overall}}$ reflects the significance of the overall biomarker--MM association; $P_{\text{nonlinear}}$ denotes evidence for departure from linearity.
    MM: multiple myeloma; HR: hazard ratio; RCS: restricted cubic spline.}
    \label{fig:spline}
}

\clearpage
    {\centering
    \includegraphics[width=\textwidth]{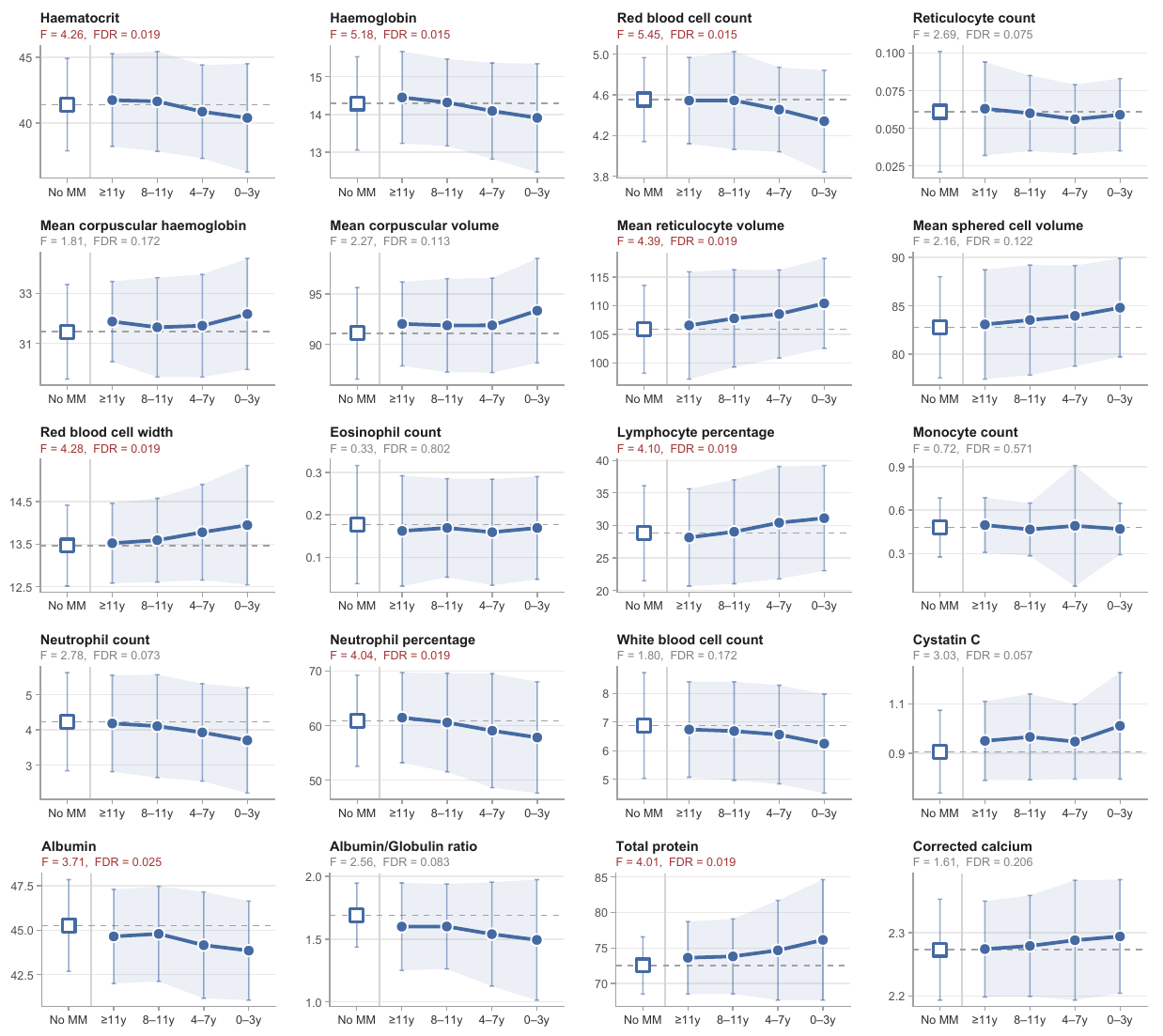}
    \captionof{figure}{\textbf{Preclinical trajectories of blood biomarkers stratified by time to multiple myeloma diagnosis.}
    Mean biomarker levels are shown across five groups: participants who did not develop MM (No MM) and incident MM cases stratified by time from baseline to diagnosis ($\geq$11 years, 8--11 years, 4--7 years, and 0--3 years before diagnosis). Error bars represent standard errors of the mean. For each biomarker, one-way ANOVA was performed across the four MM trajectory groups (excluding the No MM reference), with FDR correction applied across all biomarkers; $F$-statistics and FDR-adjusted $P$-values are shown in each panel. A monotonic trend test was additionally conducted using ordered linear regression with a group score coding of 0 (No MM) through 4 (0--3 years before diagnosis), adjusted for age and sex.
    MM: multiple myeloma; FDR: false discovery rate.}
    \label{fig:trajectory}
}

\clearpage
    {\centering
\includegraphics[width=\textwidth]{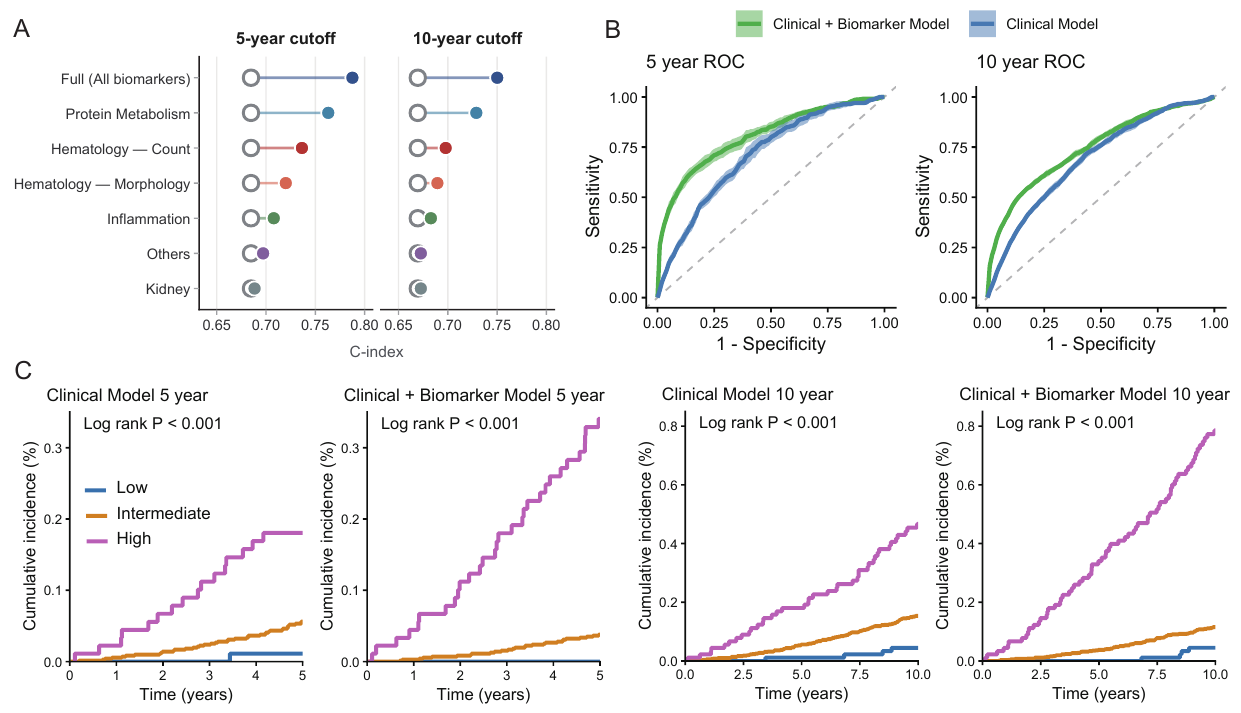}
\captionof{figure}{\textbf{Incremental value of blood biomarkers for MM risk prediction.}
\textbf{(A)} Dot plot comparing C-indices (mean $\pm$ 95\% CI across 10 cross-validation iterations) for the clinical model alone (open circles) and category-specific biomarker-augmented models (filled circles) at 5-year and 10-year prediction horizons. Biomarker categories are colour-coded; the full model incorporating all 20 biomarkers is shown at the top.
\textbf{(B)} Time-dependent receiver operating characteristic (ROC) curves at 5-year (left) and 10-year (right) horizons for the clinical model (blue) and the clinical + biomarker model (green). Shaded bands indicate 95\% confidence intervals estimated by inverse probability of censoring weighting (IPCW).
\textbf{(C)} Cumulative MM incidence curves stratified by predicted risk group (low: bottom 10\%; intermediate: 10th--90th percentile; high: top 10\%) for the clinical model (left) and clinical + biomarker model (right) at 5-year (top) and 10-year (bottom) horizons. All log-rank $P < 0.001$.
MM: multiple myeloma; ROC: receiver operating characteristic; AUC: area under the curve; IPCW: inverse probability of censoring weighting; C-index: concordance index.}
\label{fig:prediction}
}

%\clearpage
%{\centering
%\includegraphics[width=\textwidth]{Figure5S5.pdf}
%\captionof{figure}{\textbf{SHAP-based feature importance of biomarkers in the clinical + biomarker prediction model.}
%\textbf{(A)} Beeswarm plot of SHAP values (log-hazard scale) for each biomarker in the clinical + biomarker model. Each point represents one participant; colour indicates the normalised feature value from low (blue) to high (red). Biomarkers are ranked by mean $|$SHAP$|$ in descending order.
%\textbf{(B)} Bar plot of mean $|$SHAP$|$ values across all participants, summarising the average absolute contribution of each biomarker to individual MM risk predictions.
%SHAP: SHapley Additive exPlanations; MM: multiple myeloma.}
%\label{fig:shap}
%}

\end{document}